\documentclass[a4paper,12pt]{article}
\usepackage{head,fullpage}     
\usepackage{graphicx}

\usepackage{amsmath,amsfonts,amsthm,amssymb}

\usepackage[colorlinks,linkcolor=blue,citecolor=blue]{hyperref}

\usepackage{breqn}
\usepackage{titlesec}
\usepackage{enumerate}
\usepackage{tikz}
\usepackage{subcaption}
\usepackage[labelfont=bf]{caption}
\usepackage{xfrac}
\usepackage{cite}
\usepackage[titletoc]{appendix}

\titleformat*{\section}{\Large\bfseries}
\titleformat*{\subsection}{\large\itshape}
\titleformat*{\subsubsection}{\itshape}

\newcommand{\del}{\partial}
\newcommand{\C}{\mathcal{C}}
\renewcommand{\Re}{\text{Re}\,}
\renewcommand{\Im}{\text{Im}\,}
\newcommand{\Ordo}[1]{\mathcal{O}\left(#1\right)}
\newcommand{\liminfty}[1]{\lim\limits_{#1\to\infty}}

\begin{document}	
	
	{\noindent\large\bf Exact spectral solution of two interacting run-and-tumble particles on a ring lattice}
	

	\begin{minipage}{0.15\textwidth}
	\end{minipage}\hfill
	\begin{minipage}{0.85\textwidth}
			\vspace{8mm}
			\textbf{Emil Mallmin, Richard A Blythe, and Martin R Evans}
			\vspace*{2mm}\newline
			\textit{SUPA, School of Physics and Astronomy, University of Edinburgh,\\ Peter Guthrie Tait Road, Edinburgh EH9 3FD, United Kingdom}
			\vspace*{2mm}\newline
			{\small e-mail: \texttt{emil.mallmin@ed.ac.uk} }
			
			\vspace*{4mm}
			Original version October 1, 2018; updated \today
			
			\vspace{12mm}
			\textbf{Abstract}\newline
			Exact solutions of interacting random walk models, such as 1D lattice gases, offer precise insight into the origin of nonequilibrium phenomena. Here, we study a model of run-and-tumble particles on a ring lattice interacting via hardcore exclusion. 	
			We present the exact solution for one and two particles using a generating function technique. For two particles, the eigenvectors and eigenvalues are explicitly expressed using two parameters reminiscent of Bethe roots, whose numerical values are determined by polynomial equations which we derive. The spectrum depends in a complicated way on the ratio of direction reversal rate to lattice jump rate, $\omega$. For both one and two particles, the spectrum consists of separate real bands for large $\omega$, which mix and become complex-valued for small $\omega$. At exceptional values of $\omega$, two or more eigenvalues coalesce such that the Markov matrix is non-diagonalizable. A consequence of this intricate parameter dependence is the appearance of dynamical transitions: non-analytic minima in the longest relaxation times as functions of $\omega$ (for a given lattice size). Exceptional points are theoretically and experimentally relevant in, e.g., open quantum systems and multichannel scattering. We propose that the phenomenon should be a ubiquitous feature of classical nonequilibrium models as well, and of relevance to physical observables in this context.

	\end{minipage}
	
	\vfill
	
	{Submitted to \textit{J.\ Stat.\ Mech.}

	\thispagestyle{empty}	
	\newpage

	\setcounter{page}{0}                            
	\headruleheight{1pt}
	\lhead{\it Exact spectral solution of two interacting RTPs on a ring lattice}
	\rhead{\thepage}
	%
	\pagenumbering{roman}
	\tableofcontents                                
	\newpage
	\pagenumbering{arabic}

	\section{Introduction}
	Over a century ago, Einstein \cite{Einstein1905} and Smoluchowski \cite{Smoluchowski1906} explained the erratic motion of a colloidal particle as a random walk consistent with thermal equilibrium, thereby vindicating the atomic view of matter. In a modern continuation of this approach, systems out of equilibrium are modelled as random walks that break detailed balance---the equilibrium condition which must hold if the underlying microscopic dynamics of the bath is time-reversible. A single `particle' can, for instance, be out of equilibrium by being externally driven\cite{Crooks2000,Klages2013}; by being self-propelled (`active') through a mechanism not modelled in microscopic detail \cite{Bechinger2016,Fodor2018}; or by being immersed in a bath which itself consists of particles out of equilibrium \cite{Steffenoni2016}. Interacting particle systems out of equilibrium generate novel collective phenomena, including clustering \cite{Cates2015}, collective directed motion \cite{Bricard2013}, and nonequilibrium phase transitions \cite{Hinrichsen2000,Blythe2003}. Exactly solvable models have been crucial in understanding the origin of such effects \cite{Evans2002}, especially  driven lattice-gas models \cite{Schmittmann1995}, e.g.\ exclusion processes in 1D \cite{Derrida1998,Golinelli2006,Chou2011}.

	In this article we describe a continuous-time model of $N$ run-and-tumble particles on a 1D ring lattice of length $L$, interacting via hardcore exclusion, and solve the time-dependent problem exactly for one and two particles. The name `run-and-tumble particle' (RTP) is a reference to the motility patterns of bacteria such as \textit{E.\ coli}, which move in straight runs interspersed by tumble events where a new direction is chosen more or less at random \cite{Berg2004}. An RTP is more generally an example of a persistent random walker (PRW) \cite{Masoliver1989,Masoliver2017,Weiss1994}, where `persistence' refers to the correlation of successive movements induced by their dependence on an internal state. For an RTP, the internal state is the orientation $\sigma$, which in 1D takes the value $+1$ for a particle orientated to the right (i.e.\ clockwise on the ring) or $-1$ for the left (counter-clockwise). In any infinitesimal time interval $dt$, each particle has a probability $\gamma dt$ of jumping one lattice site in its current direction $\sigma$, unless the arrival site is already occupied (hardcore exclusion). This creates the possibility of particle jamming: two opposed-velocity particles on neighbouring sites blocking each other's movement. The orientation for any given particle reverses by another Poisson process with rate $\omega$, independent of any other aspects of the system. 
	The model is illustrated in Figure \ref{fig:model_def}. PRWs have been used to model photon propagation in various media \cite{Boguna1999,Miri2003}, animal movements \cite{Wu2000}, polymer conformation \cite{Fujita1980}, and molecular cargo transport \cite{Bhat2013}. The single RTP has been studied under various boundary conditions and for inhomogeneous rates \cite{Schnitzer1993,Angelani2017,Thompson2011}, and interacting RTPs at the level of fluctuating hydrodynamics \cite{Tailleur2008,Tailleur2009,Escaff2018,Thompson2011}, as well as on-lattice \cite{Thompson2011,Soto2014,Slowman2016a,Slowman2017}. Very recently, several authors have also considered the combined effects of thermal diffusion and persistent motion \cite{Demaerel2018,Malakar2018,Pietzonka2018}.
	
	\begin{figure}
		\centering
		\includegraphics[scale=0.65]{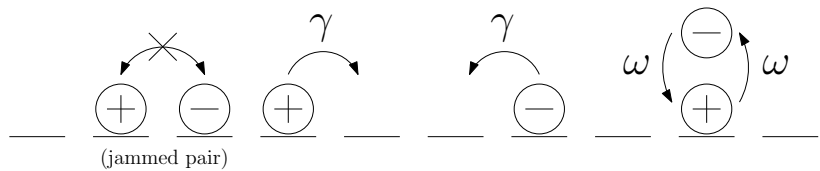}
		\caption{Transition dynamics of the RTP model with hardcore exclusion}\label{fig:model_def}
	\end{figure}
	
	Our exact two-particle solution in the centre-of-mass frame extends the result by Slowman et\ al.\ for the nonequilibrium steady state (NESS) for two particles \cite{Slowman2016a,Slowman2017}. The NESS has a non-trivial structure which puts an anomalous positive weight on the jammed configurations. We have generalized the generating function technique used in deriving the NESS to also determine all (right) eigenvectors and eigenvalues of the Markov matrix of the process. Specifically, the eigenvectors and eigenvalues are given as explicit functions of two parameters (`roots'), in turn implicitly determined by two polynomial equations which we derive. As already hinted at by the structure of the NESS, the full spectral solution bears some resemblance to the Bethe ansatz \cite{Karbach1998} which solves the related asymmetric exclusion process (ASEP) \cite{Golinelli2006}.
	
	We also detail the solution of the much simpler one-particle case. Its steady state is trivial (uniform) due to the fact that---in contrast to the two-particle case---it is kinematically reversible (discussed in Section \ref{sec:kinem}). However, since it nonetheless violates detailed balance, there is the possibility of nonequilibrium effects in the transient regime. Indeed, we find for both one and two particles that the spectrum depends in a non-trivial way on $\omega$, and is characterized by many eigenvalue crossings. These effects conspire to create non-analytic minima of the longest relaxation time of the system as a function of $\omega$, which we refer to as a dynamical transition. At a critical $\omega = \omega^*(L)$ corresponding to a non-analyticity (for one particle there is one; for two particles there a two) the relaxation time scales with system size as $\sim L$, in contrast to $\sim L^2$ for fixed values of $\omega$, including the limits $\omega \to 0$ or $\infty$. Similar scaling regimes have been found for the mixing of a related Markov chain model \cite{Diaconis2000}.
	
	The intricate $\omega$-dependence of the spectrum is a nonequilibrium phenomenon. Kato \cite{Kato1966} used the term `exceptional points' for points in parameter space where eigenvalues of a linear operator coalesce. If the spectrum of a symmetric linear operator depends on a real parameter, it will typically not have exceptional points due to the phenomenon of eigenvalue repulsion or avoided crossings \cite{Heiss1990}. Markov matrices of equilibrium models are always symmetrizable \cite[eq.\ V.6.15]{vanKampen2007}, have real spectra, and hence avoid crossings; a conclusion which does not in general hold for a nonequilibrium model. The present model interpolates between two qualitatively different situations. It becomes the symmetric simple exclusion process (SSEP) for $\omega\to \infty$, and a `quenched disordered' totally asymmetric exclusion process (TASEP) for $\omega \to 0$. On a ring lattice, the former is an equilibrium model since it satisfies detailed balance. The latter, on the other hand, is a nonequilibrium model, and in a sense strongly so, since no single transition can be reversed in a single event. Therefore, for intermediate values of $\omega$ one can expect eigenvalue crossings and real-complex transition of eigenvalues. For one- and two particles, we find that the eigenvalues come in separated real bands for $\omega$ larger than $1$ or $2$, respectively, which subsequently meet and undergo real-complex transitions for smaller $\omega$. We find exceptional points that are eigenvalue branch points and therefore correspond to non-diagonalizable eigensubspaces \cite{Kato1966}. Such exceptional points have been studied in, among other areas of physics, open quantum system and multichannel scattering where their effects can be observed experimentally \cite{Heiss2012}. It has been suggested \cite{Ryu2015} that exceptional points may be important in nonequilibrium statistical mechanics also. RTPs now constitute a concrete example.

	The article is organised as follows. In Section \ref{sec:model} the $N$-particle master equation \eqref{sec:model} is stated explicitly, and its relation to kinetic reversibility is discussed. The one-particle case is treated in Section \ref{sec:N1}, intended primarily as a prototype for and comparison to the more demanding two-particle case. Section \ref{sec:N2}, concerning the two-particle case, contains the bulk of the work presented in this article. The first subsection, Section \ref{sec:problem}, sets up the problem. Section \ref{sec:N2_outline} outlines the generating function method and pole cancellation procedure used to obtain the spectral solution in terms of the above mentioned roots. Section \ref{sec:N2_spectrum} carries out explicitly what was schematically described in the previous subsection, and can be skipped if the technical details are of less interest to the reader. The derived spectrum and eigenvectors are summarized in Section \ref{sec:N2_spectrum} and \ref{sec:N2_eigenvectors}, respectively. The long-time relaxation behaviour and dynamical transition is analyzed in Section \ref{sec:N2_relax}.	The article ends with a summary and discussion in Section \ref{sec:sum}.

	\section{Description of the $N$ run-and-tumble particle model}\label{sec:model}
	\subsection{Master equation}
	Each of the $N\ (< L)$ run-and-tumble particles is the sole occupant of one of the sites $1,2,\ldots,L$ of the ring lattice. The total system configuration is $\C = (\boldsymbol{\sigma}, \boldsymbol{n})$, with the vector of particle orientations $\boldsymbol{\sigma} = (\sigma_1,\ldots,\sigma_N)$, $\sigma_i \in\{-1,+1\}$, and the vector of particle positions $\boldsymbol{n} = \sum_{i=1}^N n_i \boldsymbol{e}_i = (n_1,\ldots,n_N)$, $n_i \in \{1,2,\ldots,L\}$. The joint probability distribution of site and orientation of every particle, $P(\C,t) \equiv P_{\boldsymbol\sigma}(\boldsymbol n, t)$, is described by the master equation to be stated below. In practice, we implement dynamics on a ring by extending the configuration space to allow $n_i \in \mathbb{Z}$, but impose a periodicity condition on the probability distribution by
	\begin{equation}\label{eq:N_period}
		P_{\sigma_1\ldots\sigma_N}(n_1, \ldots, n_N) = 	P_{\sigma_2\ldots\sigma_N\sigma_1}(n_2, \ldots, n_N, n_1 + L).
	\end{equation}
	The particle labelling convention is that the arguments appear in the order from leftmost to rightmost particle, so that relevant configurations have $1\leq n_1 < n_2 <\ldots <n_N \leq L$. When the rightmost particle jumps right from site $L$, it ends up at site $1$ and is now the leftmost particle, etc. The particles are identical but can be tracked through time and distinguished if the complete sequence of jumps and orientation flips is given. We introduce some notation to express the master equation compactly. Let $\theta_i$ be an operator acting on the $i$th particle by flipping its orientation, $\theta_i (\sigma_1,\ldots,\sigma_i,\ldots,\sigma_N) = (\sigma_1,\ldots,-\sigma_i,\ldots,\sigma_N)$; define $I_L(n,m) = \{ 1 \text{ if } n \not\equiv m \text{ mod } L; \text{ else } 0\}$ as the indicator function for $n$ and $m$ being distinct lattice sites; and allow the indexing convention $n_{- i} = n_{L-i}$. Then the master equation can be written
	\begin{dmath}\label{eq:N_ME}
	\del_t P_{\boldsymbol{\sigma}}(\boldsymbol{n}) =  \gamma \sum_{i=1}^N \left[ P_{\boldsymbol{\sigma}}(\boldsymbol{n}-\sigma_i \boldsymbol{e}_i) I_L(n_i-\sigma_i,n_{i-\sigma_i})  - P_{\boldsymbol{\sigma}}(\boldsymbol{n}) I_L(n_i+\sigma_i,n_{i+\sigma_i}) \right] 
	+\ {} \omega \sum_{i=1}^N [ P_{\theta_i \boldsymbol{\sigma}}(\boldsymbol{n}) - P_{\boldsymbol{\sigma}}(\boldsymbol{n})].
	\end{dmath}
	Without loss of generality, we henceforth choose units of time in which $\gamma = 1$. The initial condition we leave unspecified, but typically we have in mind a probability mass concentrated on a single configuration.
	
	\subsection{Lack of (kinematic) reversibility}\label{sec:kinem}

	\begin{figure}
			\centering
			\includegraphics[scale=0.65]{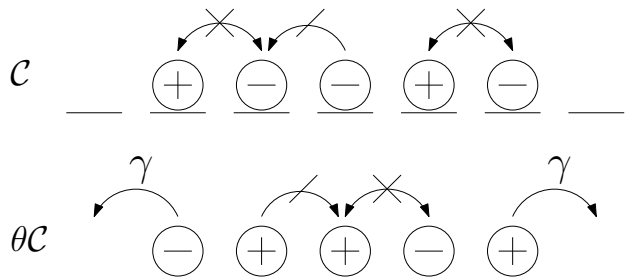}
			\caption{The configuration $\C$ has two pairs of jammed particles (and additionally one obstructed particle), whereas the kinematically reversed configuration $\theta \C$ has only one jammed pair. Consequently the difference in escape rate is $2\gamma$.}\label{fig:escape}
	\end{figure}

	The run-and-tumble model trivially breaks detailed balance since the $i$th particle, if unobstructed, may jump $n_i \to n_i + \sigma_i$, whereas the transition $n_i + \sigma_i \to n_i$ is always forbidden due to the orientation of the particle. Therefore the model is not reversible in the usual stochastic process terminology. However, once we interpret the orientations as normalized velocities, we can further ask if the steady state is statistically invariant under kinematic time-reversal of the dynamics. Let $\C$ denote a configuration of the system, and $\theta \C = (\theta_1\cdots\theta_N \boldsymbol{\sigma},\boldsymbol{n})$ its kinematic time reversal. By definition, kinematic reversibility is the requirement that
	\begin{subequations}
	\begin{gather}
	P^*(\C) = P^*(\theta \C), \label{eq:PstarSym}\\
	 P^*(\C, 0;\, \C', t) =  P^*(\theta\C', 0;\, \theta\C,t).\label{eq:Pstar2}
	\end{gather}	
	\end{subequations}
	These are the one- and two-time probability distributions in the steady state, invariant under time translations; higher order probabilities need not be considered due to the Markov property. Kinematic reversibility implies a symmetry of the Markov matrix which boils down to
	\begin{subequations}
	\begin{gather}
	\xi(\C) = \xi(\theta \C),\label{eq:Xicond}\\
	P^*(\C) W(\C\to\C') = P^*(\theta \C') W(\theta \C' \to \theta \C).\label{eq:ExtDB}
	\end{gather}
	\end{subequations}
	The first condition states the kinematic reversal symmetry of the escape rates $\xi(\C) = \sum_{\C'} W(\C \to \C')$; the second condition has been called extended detailed balance \cite{vanKampen2007}. Conditions \eqref{eq:Xicond} and \eqref{eq:ExtDB} together with \eqref{eq:PstarSym} are equivalent to kinematic reversibility (i.e.\ to \eqref{eq:PstarSym} and \eqref{eq:Pstar2}) \cite{Kelly1979}, and if they hold, the steady state distribution is Boltzmann-like, by a derivation analogous to that in the case of ordinary detailed balance. We therefore emphasize that breaking kinematic reversibility (or any analogous generalized reversibility) is necessary in order to obtain a nontrivial nonequilibrium steady state.
	
	Presently, condition \eqref{eq:Xicond} is violated unless $N=1$, as we now demonstrate. The escape rate $\xi(\C)$ equals $N\omega$ (since every particle is free to tumble) plus a $\gamma$ for every particle free to hop. Therefore, $\xi(\C) - \xi(\theta \C)$ gives the difference in unobstructed particles between the two configurations. In fact one can prove
	\begin{equation}
	\xi(\mathcal{C}) - \xi(\theta\mathcal{C}) = 2 \gamma \times \text{difference in \# of jammed particle pairs between $\theta\C$ and $\C$},
	\end{equation}
	for which Figure \ref{fig:escape} provides the intuition. Obviously, for $N = 1$ with periodic boundary conditions there can be no jamming (in contrast to the case of confining walls). Then the escape rates are identical, and due to left/right symmetry the steady state must be uniform over all states. Hence, all conditions of kinematic reversibility are satisfied. For $N > 1$, however, there always exist configurations where kinematic reversal alters the number of jammed particle pairs. Quite generally, active matter particles break kinematic reversibility due to jamming interactions with boundaries, fixed obstacles, or other particles.

	\section{Solution for one particle} \label{sec:N1}
	
	The one-particle lattice run-and-tumble model is arguably one of the simplest nonequilibrium models that depends on a parameter with a continuous range, the ratio between flipping and hopping, $\omega$. As noted above, the steady state is kinematically reversible, and in particular it is uniform. Nonetheless, because detailed balance is broken, nonequilibrium effects are manifest in the non-stationary regime. We show that the relaxation dynamics changes qualitatively at exceptional values of $\omega$ where the system is non-diagonalizable.

	\subsection{Spectral analysis}
	The objective is to obtain a spectral decomposition of the Markov matrix $M$, whose elements are defined through $\del_t P_\sigma (n) = \sum_{\sigma',n'} M_{\sigma\sigma'}(n,n') P_{\sigma'}(n')$, and analyze the spectrum's dependence on the flipping rate $\omega$. The master equations read (with $\gamma = 1$)
	\begin{dgroup}\label{eq:N1_ME}
		\begin{dmath}
		\del_t P_{+}(n,t) =  P_+(n - 1,t) - P_+(n,t) + \omega[ P_-(n,t) - P_+(n,t)],
		\end{dmath}
		\begin{dmath}
		\del_t P_{-}(n,t) = P_-(n + 1,t) - P_-(n,t) + \omega[ P_+(n,t) - P_-(n,t)],
		\end{dmath}
	\end{dgroup}
	with periodicity $P_\sigma (n + L ) = P_\sigma (n)$. They are routinely solved by introducing a Fourier transform,
	\begin{equation}
	\boldsymbol{g}(k,t) = \sum_{n = 1}^{L} z_k^n
	\begin{pmatrix}
	P_+(n, t)\\
	P_-(n,t)
	\end{pmatrix}
	,\quad z_k = \exp(2\pi i k / L), \quad k \in 1,2,\ldots,L,
	\end{equation}
	which transforms them to
	\begin{equation}\label{eq:geq}
	\del_t \boldsymbol{g}(k,t) = W(k) \boldsymbol{g}(k,t),
	\end{equation}
	where
	\begin{equation}\label{eq:Wk}
	W(k) = \begin{pmatrix}
	w(k) & \omega \\
	\omega & w(-k)
	\end{pmatrix},\quad w(k) = z_k - (1 + \omega).
	\end{equation}
	$W(k)$ corresponds to a $2 \times 2$ block on the diagonal of the Fourier transformed Markov matrix. The spectrum of $M$ is the collection of eigenvalues from all $W(k)$, and $M$ is diagonalizable if and only if $W(k)$ is diagonalizable for all $k = 1,2,\ldots L$. 
	
	\subsubsection{Spectrum: band structure and eigenvalue crossings}
	
	The eigenvalues of $W(k)$ are found to be
	\begin{equation}\label{eq:N1_spectrum}
	\lambda^{(s)}(k) = -2 \sin^2\left(\tfrac{\pi k}{L}\right) - \omega + s \sqrt{\omega^2 -  \sin^2 \left(\tfrac{2\pi k}{L}\right)},
	\end{equation}
	where $s = +1$ indicates the `right band' and $s = -1$ the `left band'. By the symmetry $\lambda(k) = \lambda(L-k)$ the spectrum is always doubly degenerate (except possibly $k = L/2$). This is clearly due to the spatial inversion symmetry of the problem. 
	 
	\begin{figure}
		\begin{subfigure}{0.5\textwidth}
			\centering			
			\includegraphics[scale=0.9]{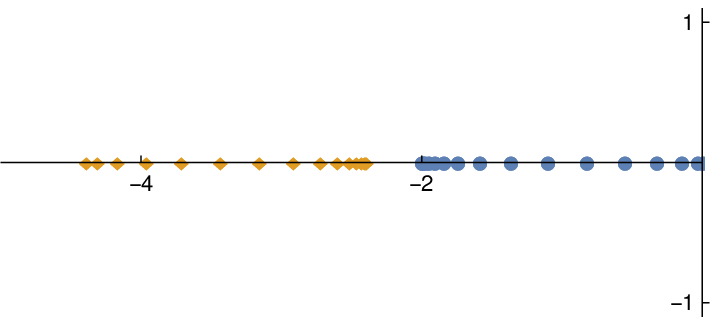}
			\caption{$\omega = 1.2$}\label{fig:N1_crossings_1}
		\end{subfigure}%
		\begin{subfigure}{0.5\textwidth}
			\centering			
			\begin{tikzpicture}
			\node[anchor=south west,inner sep=0] (image) at (0,0) 				{\includegraphics[scale=0.9]{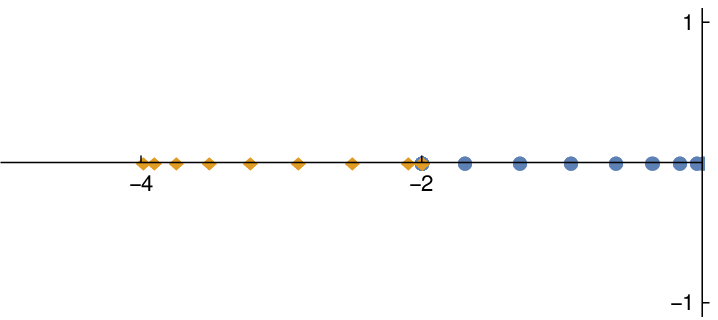}};
				\begin{scope}[x={(image.south east)},y={(image.north west)}]
					\node at (0.59,0.8) {\tiny $L$-fold degeneracy} ;
					\draw[->] (0.59,0.75) -- (0.59,0.55);
			\end{scope}
			\end{tikzpicture}		
			\caption{$\omega = 1$}\label{fig:N1_crossings_2}
		\end{subfigure}
		\begin{subfigure}{0.5\textwidth}
			\centering
			\begin{tikzpicture}
			\node[anchor=south west,inner sep=0] (image) at (0,0) 				{\includegraphics[scale=0.9]{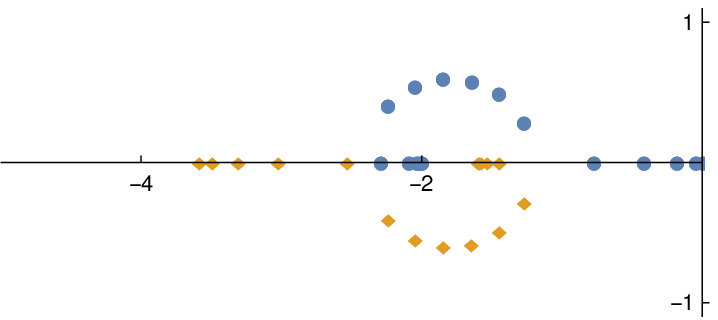}};
			\begin{scope}[x={(image.south east)},y={(image.north west)}]
			\draw[->] (0.54,0.54) -- (0.52,0.54) -- (0.52,0.6);
			\draw[->] (0.494,0.465) -- (0.52,0.465) -- (0.52,0.4);
			\draw[->] (0.705,0.465) -- (0.77,0.465) -- (0.77,0.4);
			\draw[->] (0.83,0.54) -- (0.77,0.54) -- (0.77,0.6);
			\end{scope}
			\end{tikzpicture}			
			\caption{$\omega = 0.8$}\label{fig:N1_crossings_3}		
		\end{subfigure}%
		\begin{subfigure}{0.5\textwidth}
			\centering
			\includegraphics[scale=0.9]{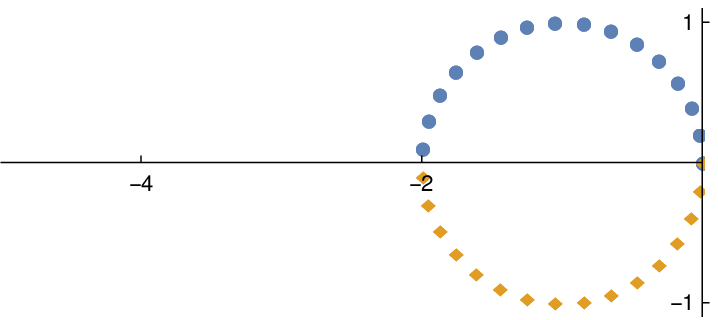}
			\caption{$\omega = 0$}\label{fig:N1_crossings_4}		
		\end{subfigure}
		\caption{One-particle spectrum \eqref{eq:N1_spectrum} in the complex plane for $L = 31$. Blue circles: right band ($s = +1$); yellow diamonds: left band ($s = -1$). (a) For $\omega > 1$ the eigenvalues come in two separate real bands. (b) At $\omega = 1$ there is an $L$-fold degeneracy at the eigenvalue $-2$. (c) As $\omega$ is further decreased, pairs of real eigenvalues cross and separate as complex conjugate pairs. The arrows indicate how the crossing is approached. (d) At $\omega = 0$ the spectrum is a unit circle shifted by $-1$.}\label{fig:N1_crossings}
	\end{figure}
	 
	We now study analytically and graphically (Figure \ref{fig:N1_crossings}) the qualitative changes to the spectrum as $\omega$ is varied. For $\omega > 1$, the spectrum \eqref{eq:N1_spectrum} is necessarily real, and for large $\omega$
	\begin{align}
	\lambda^{(+)}(k) &= - 2 \sin^2\left(\frac{\pi k}{L}\right) + \Ordo{1/\omega}, \quad
	\lambda^{(-)}(k) = - 2 \sin^2\left(\frac{\pi k}{L}\right) - 2 \omega + \Ordo{1/\omega}.
	\end{align}
	In the symmetric walk limit, $\omega \to \infty$, the left band ($s = -1$) diverges, implying the disappearance of these modes. The remaining eigenvalues are those of the symmetric walker with jump rate $1/2$. The initial jump rate $\gamma = 1$ in the direction of the particle's orientation is now split equally between left and right.
	
	At exactly $\omega = 1$, the eigenvalue $-2$ becomes $L$-degenerate (Figure \ref{fig:N1_crossings_2}) as
	\begin{equation}
	\lambda^{(s)}(k) = - 2 + \cos \left(\frac{2\pi k}{L}\right) + s \left| \cos \left(\frac{2\pi k}{L}\right) \right|.
	\end{equation}
	This degeneracy comprises the least negative half of the left band (`upper left band') and most negative half of the right band (`lower right band').
	Whereas the relaxation times of all modes are in general dependent on the system size $L$, at $\omega = 1$ half of them are not. The degeneracy allows a `macroscopic eigenvalue crossing' of the upper left and lower right bands. 
	
	For $\omega \leq 1$, at the exceptional values
	\begin{equation}\label{eq:omg_k}
	\omega_k =  |\sin(2\pi k /L)|,
	\end{equation}
	the eigenvalue $\lambda^{(-)}(k)$ from the lower (upper) left band and the eigenvalue $\lambda^{(+)}(k)$ of the lower (upper) right band coalesce. For smaller $\omega$, they leave the real line as a complex conjugate pair (Figure \ref{fig:N1_crossings_3}). Precisely at the crossing, $W(k)$, and hence also $M$, becomes non-diagonalizable. 
	The consequences for a spectral decomposition of $M$ in this case are described in Section \ref{sec:N1_proj}.

	As $\omega$ approaches zero (Figure \ref{fig:N1_crossings_4}),	
	\begin{equation}
	\lambda^{(s)}(k) = - 2 \sin^2 \left( \frac{\pi k}{L}\right) +  i s \left | \sin \left( \frac{ 2\pi k}{L}\right)\right| .
	\end{equation}
	The spectrum is identical to that of the totally asymmetric walker, defined by $\del_t P(n) = P(n - 1) - P(n)$, but four-fold degenerate as the two equivalent orientation sectors decouple.
	
	\subsubsection{The matrix exponential}\label{sec:N1_proj}
	We have derived the spectrum and now seek the projection operators onto the subspaces associated with each eigenvalue. Using them we express the matrix exponential appearing in the solution $\exp[W(k)t]\boldsymbol{g}(k,0)$ of 
	\eqref{eq:geq}. The final step of inverting the spectral solution in Fourier space to obtain the time-dependent solution of the master equations \eqref{eq:N1_ME} is omitted.
	
	The eigenvectors $\boldsymbol{u}(k)$ can be chosen $\boldsymbol{u}^{(s)}(k) = (\omega, \lambda^{(s)}(k) - w(k))^\top / \mathcal{N}^{(s)}(k)$ (for $\omega > 0$), corresponding to each eigenvalue $\lambda^{(s)}(k)$, and where $\mathcal{N}$ gives the normalization to modulus one. In the diagonalizable case when $\lambda^{(+)}(k)\neq \lambda^{(-)}(k)$ the two eigenvectors are orthogonal since $W(k)$ is symmetric, and the projection operators are simply outer products of eigenvectors. Therefore
	\begin{equation}
	\exp[W(k)t] = \sum_{s=+,-} e^{\lambda^{(s)}(k) t} \boldsymbol{u}^{(s)}(k) (\boldsymbol{u}^{(s)}(k))^\top.
	\end{equation}
	
	Less elementary is the case when $W(k)$ is non-diagonalizable. Define then $N(k) = W(k) - \lambda(k) I $. A matrix constructed in this way in general nilpotent, i.e. taken to some power becomes the zero matrix. The smallest such power will be less than or equal to algebraic multiplicity of that eigenvalue (in fact equal to the `index' of the eigenvalue; the dimension of the largest Jordan block), which presently is two. We can therefore conclude immediately that $N^2(k) = 0$, which can also be verified from the explicit expression
	\begin{equation}
	N(k) = 
	\begin{pmatrix}
	- i \sin\left(\tfrac{2\pi k}{L}\right) & \omega_k  \\
	\omega_k  &  i \sin\left(\tfrac{2\pi k}{L}\right)
	\end{pmatrix}.
	\end{equation}
	The matrix exponential then evaluates to
	\begin{equation}\label{eq:N1_nondiag_exp}
	\exp[W(k)t] = \exp[ \lambda(k)t I + N(k)t  ] = e^{\lambda(k)t}(I + N(k)t).
	\end{equation}
	That is, the dynamical modes which project onto this subspace have exponential decay, modulated by a linear time dependence.
	
	\subsection{Longest relaxation time and dynamical transition}\label{sec:N1_dyn_phase}

	From \eqref{eq:N1_spectrum}, we now determine the spectral gap, i.e., the least negative non-zero eigenvalue, which determines the longest time-scale of relaxation towards the steady state. The relaxation times are for each mode given by $\tau^{(s)}(k) = - 1/\Re \lambda^{(s)}(k)$. Clearly $\tau^{(+)}(k) \geq \tau^{(-)}(k)$, and both $\tau^{(+)}(k)$ and $\tau^{(-)}(k)$ attain their longest values for small $k$ (or very large $k$, by symmetry). The two $k = 0$ modes yield the zero eigenvalue corresponding to the steady state and a mode with relaxation time $1/(2\omega)$. This is the typical time for an initial distribution localized to one orientation sector to spread into both. We refer to this mode as the `tumble mode' since it does not contribute to the spatial relaxation dynamics: the corresponding eigenvector of $W(0)$ is $(1, - 1)^\top$ which has zero projection onto $(1, 1)$ in
	\begin{equation}\label{eq:N1_Pn_marginal}
	P(n,t)=   P_+(n,t) + P_-(n,t) = \frac{1}{L} \sum_{k=0}^{L-1} z_k^{-n}\begin{pmatrix}
	1, & 1
	\end{pmatrix}	\boldsymbol{g}(k,t). 
	\end{equation}
	We focus instead on the spatial relaxation, where the longest relaxation time is given by the $k = 1$ mode of the right band, $\tau_{\text{max}} = \tau^{(+)}(1)$. With
	\begin{equation}
	\omega^*(L) = \sin(2\pi/L),
	\end{equation}
	i.e.\ $\omega^* = \omega_1$ as in \eqref{eq:omg_k},	we find
	\begin{equation}
	1/\tau_{\text{max}} = 
	\begin{cases}
		  1 + \omega - \sqrt{1 - {\omega^*}^2 } - \sqrt{\omega^2 - {\omega^*}^2} , & \omega > \omega^* \\
		 1 + \omega - \sqrt{1 - {\omega^*}^2 }, & \omega \leq \omega^*.
	\end{cases}
	\end{equation}
	$\tau_{\text{max}}$ is plotted in Figure \ref{fig:N1_tau_max}. For any $L$, it is minimized at $\omega = \omega^*$ at which $\tau_{\text{max}}$ has a cusp and is non-analytic. (The shortest possible relaxation of the $k$th mode similarly occurs at $\omega = \omega_k$.) Since $\omega^*$ is in general small,
	\begin{equation}
	\tau_{\text{opt}} = \tau_{\text{max}}|_{\omega=\omega^*}\ \approx \frac{1}{\omega^*} \sim L.
	\end{equation}
	In comparison, for constant $\omega$, and in particular for both the symmetric and totally asymmetric walk limits, one finds the scaling $\tau_{\text{max}} \sim L^2$. It is possible to obtain a scaling $\tau_{\text{max}} \sim L^\alpha$, $1 \leq \alpha \leq 2$, by choosing $\omega \sim L^{-\alpha}$.
	
	\begin{figure}
		\centering
		\begin{tikzpicture}
		\node[anchor=south west,inner sep=0] (image) at (0,0) 				{\includegraphics[scale=0.9]{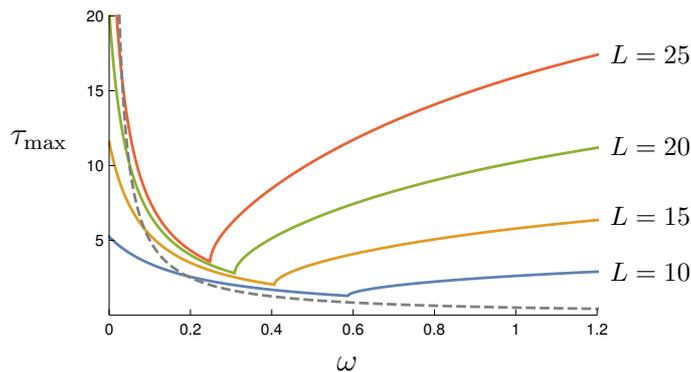}};
		\begin{scope}[x={(image.south east)},y={(image.north west)}]
		\node at (-0.09,0.6) {$\tau_{\text{max}}$};
		\node at (0.5,-0.08) {$\omega$};
		\node at (1.08, 0.86) {\footnotesize$L = 25$};
		\node at (1.08, 0.58) {\footnotesize$L = 20$};
		\node at (1.08, 0.37) {\footnotesize$L = 15$};
		\node at (1.08, 0.2) {\footnotesize$L = 10$};
		\end{scope}
		\end{tikzpicture}		
		\caption{One-particle relaxation times. Coloured lines: longest relaxation time involved in spatial dynamics. The global cusp-shaped minima occur at $\omega = \omega^*(L)$. Dashed line: the tumble mode with relaxation $1/(2\omega)$.}\label{fig:N1_tau_max}
	\end{figure}
	
	In summary, the non-analyticity at the exceptional point $\omega^*$ separates a region of strictly exponential relaxation ($\omega > \omega^*$) from one of oscillatory exponential relaxation ($\omega < \omega^*$). At exactly $\omega^*$, the relaxation is exponential modulated by a linear time dependence; cf.\ \eqref{eq:N1_nondiag_exp}. This value minimizes the relaxation time and has the optimal system size scaling. We will refer to the existence of qualitatively different dynamical regimes separated by an exceptional point as a dynamical transition. To give a physical explanation for the presence of a transition in this model, we suggest there may be a trade-off between rapid phase space exploration and probability diffusion, such that a unique relaxation optimum arises. Since the steady state is uniform over all states, every site must be visited many times before spatial relaxation has occurred. With a high degree of persistence, every state is visited in a relatively short time; if $\omega \sim 1/L$, this implies a finite probability (even for large $L$) of making $L$ jumps in the same direction before reversing, where $L$ jumps (of either direction) on average occurs on a time of scale $L$, since  $\gamma = 1$. An inverse flipping rate comparable to the system spanning time may enhance the local diffusion of probability, which is due to statistical variance in the process, without undermining persistence, which benefits from less variance. In contrast, for large flipping rates, sites far from the initial one would take a long time to reach.
	
	\section{Solution for two particles in centre-of-mass frame}\label{sec:N2}
	We now present the exact solution for two particles. In contrast to the one-particle case, a generalized detailed balance is now lacking so that a genuine nonequilibrium steady state pertains. Here, we generalize the generating function approach used in \cite{Slowman2016a,Slowman2017} to find the steady state, in order to find also the elementary solutions of the master equations, i.e.\ the set of eigenvalues and (right) eigenvectors of the Markov matrix. The mathematical problem is simplified by describing the dynamics in the centre-of-mass frame; the two position arguments are replaced by the particle-particle separation. The situation becomes similar (although not identical) to a single run-and-tumble particle in a confined space. Furthermore, the symmetries of the master equations can be exploited to yield two distinct sectors of eigenvalues and eigenvectors. Most of the complexity lies in the sector we call `symmetric'. Its eigenvalues and eigenvectors are naturally expressed in terms of two complex parameters $z_1$ and $z_2$, reminiscent of Bethe roots; see \eqref{eq:lambda_of_z1} and \eqref{eq:solution}. For given system parameters $L,\omega$, the set of root values producing the spectrum are given by the solutions to the polynomial equations \eqref{eq:polyeqs}. Several observations from the one-particle case carry over to the spectral structure of the two-particle solution. There are several regimes in the parameter $\omega$ separating qualitatively different spectra. It follows broadly the previous pattern of real branches of eigenvalues which cross and spread into the complex plane with diminishing $\omega$, but the structure is richer and not easily summarized. Again, the relaxation times have minima and cusps at exceptional values of the flipping rate. In particular, this is true of the longest relaxation time, which has two cusps, each of which indicates a dynamical transition. 	
	
	\subsection{Spectral decomposition problem}\label{sec:problem}
	Consider the joint probability of particle orientations and their relative displacement $n$,
	\begin{equation}
	P_{\sigma_1 \sigma_2}(n) \propto  \sum_{n_1 \in \mathbb{Z}} \sum_{n_2 \in \mathbb{Z}}  \delta_{n,n_2-n_1} P_{\sigma_1 \sigma_2}(n_1,n_2),
	\end{equation}
	where we consider only $1 < n < L-1$. 
	This probability satisfies the following master equation, which can be derived from \eqref{eq:N_ME} for $N=2$, or written down directly from the particle dynamics, as	
	\begin{dgroup}\label{eq:MEsep}
		\begin{dmath}\label{eq:N2_ME_++}
		\del_t P_{++}(n) =   P_{++}(n+1)I_{ n \neq L - 1} + P_{++}(n - 1) I_{  n \neq 1} 
		+ \omega \left[ P_{+-}(n) + P_{-+}(n)\right] 
		\\-  \left[ I_{  n \neq 1} + I_{  n \neq L - 1} + 2\omega\right] P_{++}(n),
		\end{dmath}
		\begin{dmath}\label{eq:N2_ME_+-}
		\del_t P_{+-}(n) =  2 P_{+-}(n+1) I_{  n \neq L - 1}
		+ \omega \left[P_{++}(n) + P_{--}(n)\right] 
		\\ - 2\left[I_{  n \neq 1} + \omega\right]P_{+-}(n),
		\end{dmath}
		\begin{dmath}\label{eq:N2_ME_-+}
		\del_t P_{-+}(n) = 2 P_{-+}(n-1) I_{  n \neq 1}
		+ \omega \left[P_{++}(n) + P_{--}(n)\right] 
		\\ - 2\left[I_{  n \neq L - 1} + \omega\right]P_{-+}(n).
		\end{dmath}
		\begin{dmath}\label{eq:N2_ME_--}
		\del_t P_{--}(n) = P_{--}(n + 1) I_{  n \neq L - 1}  + P_{--}(n-1)I_{  n \neq 1} 	+ \omega \left[ P_{+-}(n) + P_{-+}(n)\right] 
		\\ - \left[ I_{  n \neq L - 1} + I_{  n \neq 1} + 2\omega\right]P_{--}(n),
		\end{dmath}
	\end{dgroup}
	where $I_Q = \{1 \text{ if } Q \text{ is true}; \text{else } 0\}$. The periodicity constraint \eqref{eq:N_period} becomes
	\begin{equation}\label{eq:period}
	P_{\sigma_1\sigma_2}(n) = P_{\sigma_2\sigma_1}(L - n).
	\end{equation}
	The particle separation argument $n$ has been defined as a clockwise measurement from the first particle to the second, as ordered by the indices. Equation \eqref{eq:period} expresses the arbitrariness of which  particle is labelled the `first'.

	In the following, we use a generating function approach to solve the eigenvalue equations obtained by substitution of the elementary solutions 
	$P_{\sigma_1\sigma_2}(n) \propto e^{\lambda t} u_{\sigma_1\sigma_2}(n)$ into \eqref{eq:MEsep}. That is, we seek the right eigenvector $u$ with components $u_{\sigma_1\sigma_2}(n)$ corresponding to the eigenvalue $\lambda$ of the Markov matrix $M$. Since $M$ is neither symmetric nor symmetrizable\footnote{We later find numerically that in some ranges of $\omega$ it has complex eigenvalues. Therefore, it cannot always be similar to a symmetric matrix.}, a complete spectral solution to the problem \eqref{eq:MEsep} would require also the left eigenvectors. Their derivation is omitted, but could in principle be found by the same method we use for the right eigenvectors. 
	
	\subsection{Outline of the generating function method}\label{sec:N2_outline}
	\subsubsection{Eigenspace symmetrization}\label{sec:eigsym}
	Before introducing the generating function we can simplify the eigenvalue problem by exploiting the formal symmetries of the equations \eqref{eq:MEsep}. These symmetries are $++\leftrightarrow --$ and $(+-,n)\leftrightarrow(-+,L-n)$, which imply, respectively, that if $u$ is an eigenvector with eigenvalue $\lambda$, then so are $u'$ and $u''$ with components
	\begin{subequations}
		\begin{gather}
			u_{\pm \mp}'(n) = u_{\pm\mp}(n),\\ u_{\pm\pm}'(n) = u_{\mp\mp}(n),  \label{eq:sym_pprim}			
		\end{gather}
	\end{subequations}
	\begin{equation}
		u_{\sigma_1\sigma_2}''(n) = u_{\sigma_2\sigma_1}(L-n). \label{eq:sym_prim} 
	\end{equation}
	This allows the construction of symmetric ($+$) or antisymmetric ($-$)  eigenvectors. For example, $u$ and $u'$ can be combined in two independent ways as
	\begin{equation}
	u_{\sigma_1\sigma_2}(n) \pm u_{\sigma_2\sigma_1}(L-n).
	\end{equation} 
	Therefore, for an arbitrary eigenvector $u$, we may always assume
	\begin{gather}
	u_{++}(n) = r\, u_{--}(n), \label{eq:sym_r} \\
	u_{\sigma_1\sigma_2}(n) = s\, u_{\sigma_2\sigma_1}(L-n),\label{eq:sym_s}
	\end{gather} 
	where $s,r = \pm 1$. Each combination $(r,s)$ gives one of four sectors of the eigenspace, containing only eigenvectors with the prescribed symmetry. In principle, eigenvalues could be degenerate, with eigenvectors belonging to different sectors. As it turns out, in this problem the eigenvalues are generically (i.e.\ with only a few exceptions) non-degenerate, and can therefore be said to belong to a given sector. The periodicity \eqref{eq:period} implies that the probability distribution will be constructed only from eigenmodes with the symmetry $s = +1$. Thus, only about half of the eigenmodes of $M$ are relevant, and we save some work by ignoring the other half from the outset. Of the two relevant sectors, we will call $(r,s) = (+1,+1)$ the symmetric sector, and the $(-1,+1)$ the antisymmetric sector.
	
	\subsubsection{Generating function equation and pole cancellation procedure}
	For the present problem we define a vector-valued generating function by
	\begin{equation}
	\boldsymbol{g}(x) = \sum_{n=1}^{L-1} x^n \boldsymbol{u}(n),\quad \boldsymbol{u}(n) = \begin{pmatrix}
	u_{++}(n)\\
	u_{+-}(n)\\
	u_{-+}(n)
	\end{pmatrix}.
	\end{equation}
	It is invertible by $\boldsymbol{u}(n) = (1/n!) (\del_x^n \boldsymbol{g})(0).$ Because of \eqref{eq:sym_r}, it is not necessary to include $u_{--}(n)$ in $\boldsymbol{u}(n)$. Multiplying each eigenvalue equation by $x^n$  and summing over $n$, one obtains a set of linear equations in $g_{\sigma_1\sigma_2}(x)$ which can be summarized as
	\begin{equation}\label{eq:g_eq_first}
	(A(x) - \lambda I) \boldsymbol{g}(x) = \boldsymbol{b}(x),\quad \text{for } x \neq 0,
	\end{equation}
	and by definition $\boldsymbol{g}(0) = 0$. The matrix $H(x,\lambda) \equiv A(x) - \lambda I$ and vector $\boldsymbol{b}(x)$ are given explicitly in the next section, and depend on the sector $(r,s)$. For now, we only outline the logical steps in solving \eqref{eq:g_eq_first}. The elements of $A(x)$ are all rational functions of $x$, and therefore the determinant of $H(x,\lambda)$ is too. We have the two alternative factorizations
	\begin{equation}\label{eq:detH}
	\begin{split}
	\det H(x,\lambda) &= - \prod_{i=1}^3 (\lambda - \lambda_i(x))
	= \frac{1}{p(x,\lambda)} \prod_{i=1}^{m} (x - z_i(\lambda)).
	\end{split}
	\end{equation}
	The denominator $p(x,\lambda)$ is a polynomial in $x$ with zeroes different from the $m$ zeroes $z_i(\lambda)$ of the numerator. Since $m$ is finite for a given $\lambda$, and the spectrum is a finite set, $H(x,\lambda)$ will be invertible for all but a finite number of parameter values. Therefore, generically,
	\begin{equation}\label{eq:g_H_inv}
	\boldsymbol{g}(x) = \frac{p(x,\lambda)}{\prod_{i=1}^{k} (x - z_i(\lambda))} C^\top(x,\lambda)\boldsymbol{b}(x),
	\end{equation}
	where $C^\top(x,\lambda)$ is the transposed matrix of cofactors of $H(x,\lambda)$. Since $\boldsymbol{g}(x)$ is continuous, the above expression must hold even in the limits $x \to z_i(\lambda)$ for which $H(x,\lambda)$ is not invertible. For this limit to exist we require that the poles in the denominator be cancelled by zeroes of corresponding order in $C^\top \boldsymbol{b}$,
	\begin{equation}\label{eq:N2_Ctb}
	C^\top (z_i(\lambda), \lambda) \boldsymbol{b}(z_i(\lambda)) = 0,\quad \text{for } i = 1,\ldots, m.
	\end{equation}
	This constitutes a set of implicit equations in $\lambda$ that determine the spectrum in full. It will be advantageous to change basic variables from $\lambda$ to the set of `roots' $z_i = z_i(\lambda)$. To close the equations in these variables, additional constraints relating the $z_i$ are derived by eliminating $\lambda$ between $z_1 = z_1(\lambda)$, $z_2 = z_2(\lambda)$, etc.
	
	The eigenvectors will also have a clear structure in terms of these roots. Essentially, they are derived from \eqref{eq:g_H_inv} by a partial fraction decomposition leading to a sum of geometric series in $x$. The eigenvector components are then read off as the coefficients in the series expansion.
	
	\subsection{Parametrization in the roots $z_1$ and $z_2$ }\label{sec:roots}
	In this section we construct explicitly the roots $z_i$ introduced schematically in the previous section, and express the generating function in terms of them. We must treat the symmetry sectors (as introduced in Section \ref{sec:eigsym}) separately. The derivation of eigenvectors by inverting the generating function parametrized by the roots is deferred to Appendix \ref{app:eigfun}. The main result of this section is the derivation of the root-parametrized eigenvalue equations \eqref{eq:polyeqs} for the symmetric sector, and a direct derivation of the spectrum for the antisymmetric sector.
	
	For notational compactness, we introduce the shifted eigenvalue
	\begin{equation}
	\zeta = \lambda + 2(1 + \omega).\label{eq:zeta}
	\end{equation}
	Furthermore, the following two functions will be used extensively,
	\begin{equation}\label{eq:mu_nu}
	\mu(x) = x - \frac{ \zeta }{2},\quad \nu(x) = \frac{1}{x} - \frac{\zeta}{2}.
	\end{equation}
	Note the useful relation $\nu(x) = \mu(1/x)$. (In \eqref{eq:mu_nu}, and henceforth, we omit the function arguments indicating the $\lambda$-dependence.)
	
	Upon multiplying the eigenvalue equations by $x^n$ and summing over $n$ we find	
	\begin{subequations}\label{eq:geqs}
	\begin{gather}
	(\mu(x) + \nu(x))g_{++}(x) + \omega (g_{+-}(x) + g_{-+}(x)) = (1-x)(1-sx^{L-1}) u_{++}(1) \label{eq:g++}\\
	\nu(x) g_{+-}(x) + \omega \delta_{r,1} g_{++}(1) = (1-x) u_{+-}(1),\\
	\mu(x) g_{-+}(x) + \omega \delta_{r,1} g_{++}(1) = -s(1-x)x^{L-1} u_{+-}(1),\\
	(\mu(x) + \nu(x))g_{++}(x) + r\omega (g_{+-}(x) + g_{-+}(x)) = (1-x)(1-sx^{L-1}) u_{++}(1). \label{eq:g--}
	\end{gather}
	\end{subequations}
	Here, \eqref{eq:sym_r} and \eqref{eq:sym_s} have been applied. The symmetric sector is the larger and more complex, and, as will be shown, is parametrized by two roots, $z_1$ and $z_2$. The antisymmetric sector is comparatively simple and depends only on (the same) $z_1$. We treat each sector in turn, assuming $\omega > 0$, and lastly consider the limit case $\omega = 0$.	
	
	\subsubsection{The symmetric sector}
	In this sector, the matrix $H(x) = A(x) - \lambda I$ appearing in \eqref{eq:g_eq_first} becomes
	\begin{equation}\label{eq:H}
	H(x) = 
	\begin{pmatrix}
	\mu(x) + \nu(x) & \omega & \omega \\
	\omega & \nu(x) & 0 \\
	\omega & 0 & \mu(x)
	\end{pmatrix}.
	\end{equation}
	Its determinant is
	\begin{equation}
	\det H(x) = (\mu(x)+\nu(x))(\mu(x)\nu(x) - \omega^2),
	\end{equation}
	which vanishes for $x$ in the set of roots $ \{ z_1, \frac{1}{z_1}, z_2, \frac{1}{z_2}\}$, where
	\begin{equation}\label{eq:N2_z_roots_def}
	\mu(z_1) + \nu(z_1) = 0,\quad \mu(z_2)\nu(z_2) = \omega^2.
	\end{equation}
	From here on we refer only to $z_1$ and $z_2$, and not their reciprocals, as `the roots'. Solving the quadratic equations \eqref{eq:N2_z_roots_def} yields
	\begin{subequations}\label{eq:def_z1z2}
		\begin{gather}
		z_1 = \frac{\zeta}{2} + \sqrt{\frac{\zeta^2}{4} - 1}, \label{eq:z1def} \\
		z_2 = \frac{\zeta}{4} + \frac{1}{\zeta}\Big(1-\omega^2\Big) + \frac{1}{2}\sqrt{\frac{\zeta^2}{4} - 2(1+\omega^2) + \left( \frac{2(1-\omega^2)}{\zeta} \right)^2}. \label{eq:z2def}
		\end{gather}
	\end{subequations}
	(The reciprocals differ by the sign of the square root terms.) It follows that
	\begin{subequations}
		\begin{gather}
		(x - z_1)(x - 1/z_1) = x(\mu(x) + \nu(x)),\\
		(x - z_2)(x - 1/z_2) = - \frac{2x}{\zeta}( \mu(x)\nu(x) - \omega^2),
		\end{gather}		
	\end{subequations}
	whence $p(x,\lambda)$ appearing in \eqref{eq:g_H_inv} equals $-2x^2/\zeta$.
	
	At this point the eigenvalues are not yet known. Instead, we seek tractable and closed equations in $z_1$ and $z_2$, whose solutions then produce the eigenvalues through the inversion of (for example) \eqref{eq:z1def},
	\begin{equation}
	\lambda = z_1 + \frac{1}{z_1} - 2(1 + \omega).\label{eq:lambda_of_z1}
	\end{equation}
	Since there are two independent variables ($z_1$ and $z_2$) we need two
	equations, which we refer to as root-parametrized eigenvalue equations.
	
	The first one is derived by eliminating $\zeta$ between the two equations \eqref{eq:N2_z_roots_def}. Employing the notational shorthand $\bar{z} = 1/z$, the result is 
	\begin{equation}\label{eq:N2_zpoly2_first}
	(z_1 + \bar{z}_1)\left[ 2(z_2 + \bar{z}_2) - (z_1 + \bar{z}_1) \right] = 4(1-\omega^2).
	\end{equation}
	Already, it is apparent that $\omega = 1$ is a distinguished value. 
	
	The second root equation is derived from the pole cancellation condition \eqref{eq:N2_Ctb} and is more involved. We require the transposed matrix of cofactors
	\begin{equation}
	C^\top = \begin{pmatrix}
	\mu \nu & - \mu \omega & - \nu \omega \\
	- \mu \omega & \mu(\mu+\nu) - \omega^2 & \omega^2 \\
	-\nu \omega & \omega^2 & \nu(\mu+\nu) - \omega^2
	\end{pmatrix}
	\end{equation}
	and the vector $\boldsymbol{b}(x)$ which we decompose as
	\begin{equation}
	\boldsymbol{b}(x) = B(x)\tilde{\boldsymbol{u}},
	\end{equation}
	where  
	\begin{equation}
	B(x) = (1 - x) \begin{pmatrix}
	1 - x^{L-1} & 0 & 0 \\0 & 1 & 0 \\
	0 & 0 & - x^{L-1}
	\end{pmatrix},
	\end{equation}
	and\footnotemark
	\begin{equation}
	\tilde{\boldsymbol{u}} = u_{++}(1) \begin{pmatrix}
	1 \\ 0 \\ 0
	\end{pmatrix}
	+ u_{+-}(1) \begin{pmatrix}
	0 \\ 1 \\ 1
	\end{pmatrix}.
	\end{equation}
	\footnotetext{The fact that $\boldsymbol{b}(x)$ depends through $\tilde{\boldsymbol{u}}$ on the components of $(\del_x \boldsymbol{g})(0)$  implies no additional constraints on $u_{++}(1)$ and $u_{+-}(1)$ over $\tilde{\boldsymbol{u}} \neq 0$ and whatever comes of the pole cancellation condition; the inversion \eqref{eq:g_H_inv} is already self-consistent for $(\del_x \boldsymbol{g})(0)$.}
	The pole cancellation condition \eqref{eq:N2_Ctb} gives one equation per component, of which there are three, for each of the four poles $z_1, \frac{1}{z_1}, z_2, \frac{1}{z_2}$. However, due to the reciprocity of the poles, and the relations \eqref{eq:N2_z_roots_def}, they are not independent. One finds that all the conditions are satisfied if and only if 
	\begin{dgroup}\label{eq:Cb_zero}
		\begin{dmath}\label{eq:Cb_zero_1}
		(1-z_1) [ \omega \mu(z_1) (1-z_1^{L-1})u_{++}(1) + \omega^2 (1+z_1^{L-1})u_{+-}(1) ] = 0,
		\end{dmath}
		\begin{dmath}\label{eq:Cb_zero_2}
		(1-z_2) \mu(z_2)[ - \omega (1-z_2^{L-1}) u_{++}(1) + (\mu(z_2) - \nu(z_2)z_2^{L-1})u_{+-}(1)] = 0.
		\end{dmath}
	\end{dgroup}
	We think of $u_{++}(1)$ and $u_{+-}(1)$ as constants (for a given $\lambda$), so at first sight \eqref{eq:Cb_zero_1} and \eqref{eq:Cb_zero_2} may appear to be independent. However, they are not, as becomes clear from writing them in matrix form as
	\begin{equation}\label{eq:N2_pole_cancellation_matrix}
	\begin{pmatrix}
	1 - z_1 & 0 \\0 & 1-z_2
	\end{pmatrix}	
	\begin{pmatrix}
	\mu(z_1) (1 - z_1^{L-1} ) & \omega (1 + z_1^{L-1}) \\
	-\omega (1 - z_2^{L-1}) & \mu(z_2) - \nu(z_2)z_2^{L-1} 
	\end{pmatrix}
	\begin{pmatrix}
	u_{++}(1) \\
	u_{+-}(1)
	\end{pmatrix}
	=
	\begin{pmatrix}
	0 \\
	0
	\end{pmatrix}.
	\end{equation}	
	We require $\tilde{\boldsymbol u} \neq \boldsymbol{0}$ as otherwise $u = 0$ and is not an eigenvector. Therefore, the determinant of the matrix product in \eqref{eq:N2_pole_cancellation_matrix} must vanish. This implies the two possibilities
	\begin{subequations}\label{eq:pcconds}
	\begin{gather}
	(1-z_1)(1-z_2) = 0 \label{eq:pre_Jdet0}\\ \text{or}\nonumber\\	
	\mu(z_1)(1 - z_1^{L-1}) (\mu(z_2) - \nu(z_2)z_2^{L-1}) = - \omega^2 (1+z_1^{L-1})(1-z_2^{L-1}), \label{eq:pre_Jdet}
	\end{gather}
	\end{subequations}
	from which we will derive the second root equation.
	
	If \eqref{eq:pre_Jdet0} is satisfied, then either $z_1=1$ or $z_2 = 1$, which creates a double pole in \eqref{eq:g_H_inv}. The other possibilities of having double poles come from $z_1 = -1$, or $z_2 = -1$, or $z_1 = z_2$, which are in fact possible solutions of \eqref{eq:pre_Jdet}. In the symmetric sector, only $z_2 = 1$ cancels the double pole by a corresponding double zero in \eqref{eq:Cb_zero}. We leave the treatment of this special case to the end of this section. $z_1 = 1$ is admissible in the asymmetric sector, treated in the next section. The other cases are inconsistent. There are further special solutions of \eqref{eq:pcconds} for which a factor on either side of the equality  \eqref{eq:pre_Jdet} evaluates to zero, e.g.\ $z_1^{L-1} = z_2^{L-1} = 1$. These cases confine both roots to a discrete set of possible values independently of $\omega$. However, the first root equation \eqref{eq:N2_zpoly2_first} must also be satisfied and it depends explicitly on $\omega$. Therefore, these cases can only be consistent for particular fine-tuned values of $\omega$. We omit the derivation of these values, noting only that they occur for $\omega \leq 2$ and are expected to be the exceptional values for which eigenvalue crossings occur, and to imply non-diagonalizability; cf.\ the one-particle analysis. \newline 
	
	\noindent\textit{Generic case.}\quad Leaving these special cases for now, we continue with the generic case in which \eqref{eq:pre_Jdet} is non-trivially satisfied. Then solving for $u^{++}(1)/u^{+-}(1)$ in \eqref{eq:Cb_zero_1} and \eqref{eq:Cb_zero_2} we get
	\begin{equation}\label{eq:eigeneq1}
	\frac{u^{++}(1)}{u^{+-}(1)} = -\frac{\omega}{\mu(z_1)}\frac{1 + z_1^{L-1}}{1 - z_1^{L-1}} = \frac{\mu(z_2) - \nu(z_2)z_2^{L-1}}{\omega(1 - z_2^{L-1})}.
	\end{equation}	
	Making judicious use of \eqref{eq:N2_z_roots_def}, we can rewrite the right equality of \eqref{eq:eigeneq1} explicitly in terms $z_1$ and $z_2$, arriving at our second root-parameterized eigenvalue equation \eqref{eq:zpoly1}, stated besides the first, \eqref{eq:N2_zpoly2_first},
	\begin{subequations}\label{eq:polyeqs}
		\begin{gather} 
		(z_1 + \bar{z}_1)\left[ 2(z_2 + \bar{z}_2) - (z_1 + \bar{z}_1) \right] = 4(1-\omega^2), \label{eq:zpoly2}\\		
		z_2^{L-1} = \frac{2 z_2 - (z_1 + \bar{z}_1)}{2 \bar{z}_2 - (z_1 + \bar{z}_1)}\cdot \frac{(\bar{z}_1 - \bar{z}_2) + (z_1 - \bar{z}_2)z_1^{L-1}}{(\bar{z}_1 - z_2) + (z_1 - z_2)z_1^{L-1}},\label{eq:zpoly1}
		\end{gather}
	\end{subequations}
	where, as before, $\bar{z} =1/z $. The equations \eqref{eq:polyeqs} furnish the exact solution of the spectrum in the symmetric sector (excepting the special cases listed previously). Note that both equations are invariant under either of the transforms $z_1 \to \bar{z}_1$ or $z_2 \to \bar{z}_2$. Hence, if $(z_1,z_2)$ is a solution, then so are $(z_1,\bar{z}_2)$,  $(\bar{z}_1,z_2)$, and $(\bar{z}_1,\bar{z}_2)$. Nonetheless, they give the same eigenvalue. Note also that $z_1 = z_2$ remains a spurious solution of \eqref{eq:polyeqs}, as explained above, and must be discarded.

	\noindent\textit{Special case $z_2 = 1$.}\quad We return now this case, responsible for two special modes: the steady state and a `tumble mode' with the $L$-independent eigenvalue $\lambda = -4\omega$, analogous to the one-particle case. Solving \eqref{eq:N2_zpoly2_first}, the steady state has the real root
	\begin{equation}
	z_1 = 1 + \omega + \sqrt{\omega(2+\omega)}
	\end{equation}
	(cf.\ \cite{Slowman2016a}). The tumble mode has
	\begin{equation}
		z_1 = 1-\omega + \sqrt{\omega(\omega - 2)},
	\end{equation}
	which transitions from the real line to the unit circle for $\omega < 2$. In both cases, in order to ensure $\tilde{\boldsymbol u} \neq \boldsymbol{0}$ we must choose from \eqref{eq:Cb_zero_1}
	\begin{equation}\label{eq:u_frac_S1}
	\frac{u_{++}(1)}{u_{+-}(1)} = \frac{\omega}{\mu(z_1)} \frac{1 + z_1^{L-1}}{1 - z_1^{L-1}}.
	\end{equation}
	(This is the left equality of \eqref{eq:eigeneq1} and $z_2 = 1$ also solves \eqref{eq:polyeqs}. In practice, we will not need to treat this case separately from the generic one when deriving the eigenvectors or solving \eqref{eq:polyeqs} numerically.) For the tumble mode, \eqref{eq:u_frac_S1} is undefined for $\omega$ such that $z_1$ is an $(L-1)$-root of 1. Such $\omega$ correspond to an eigenvalue crossing with a mode from the antisymmetric sector, as becomes clear from the next section.
			
	\subsubsection{The antisymmetric sector}
	In this sector, the set of equations \eqref{eq:geqs} immediately reduces to	\begin{subequations}
		\begin{gather}
		g_{++}(x) = \frac{x(1-x)(1-x^{L-1})}{(x-z_1)(x-1/z_1)} u_{++}(1),\label{eq:g++asym}\\
		g_{+-}(x) = g_{-+}(x) = 0,
		\end{gather}
	\end{subequations}
	with $z_1$ defined by \eqref{eq:z1def} as before. The poles can be cancelled if and only if $z_1^{L-1} = 1$. Letting $\theta_m = m\pi/(L-1)$, $m = 1,2,\ldots L-1$, the eigenvalues are
	\begin{equation}
	\lambda_m = - 4 \sin^2 \theta_m - 2\omega.\label{eq:lambda_asym}
	\end{equation}
	For the case of $z_1 = 1$, the eigenvalue is $\lambda = -2\omega$, which we denote as another $L$-independent tumble mode. The relaxation time of this mode is $1/(2\omega)$, twice that of $1/(4\omega)$ for the tumble mode in the symmetric sector. If the probability distribution has an initial condition with $P_{++}(n,0) = P_{--}(n,0)$ it will not involve the antisymmetric sector, and hence not the slower decaying tumble mode. We may therefore suppose that it is related to the spreading of the probability mass between orientation sectors where both particle move in the same direction (i.e.\ between the $++$ and $--$ sector). Similarly, the faster tumble mode should be related to relaxation between same-direction and opposite-direction orientation sectors (i.e.\ between either of $+-$/$-+$ and either of $++$/$--$).

	\subsubsection{The $\omega = 0$ limit}
	For $\omega = 0$ the orientation sectors are decoupled (i.e.\ the Markov matrix is completely reducible). Now $r$ disappears from the equations \eqref{eq:geqs}, indicating a double degeneracy of the eigenvalues. 	
	Remaining in $s=+1$, $g_{++}(x)$ is identical to \eqref{eq:g++asym}, whereas	
		\begin{gather}	
		g_{+-}(x) = \frac{x(1-x)}{1 - \frac{\zeta}{2} x} u_{+-}(1),\quad
		g_{-+}(x) = \frac{(1-x)x^{L-1}}{\frac{\zeta}{2} - x} u_{+-}(1).
		\end{gather}
	If $u_{+-}(1)$ is to be non-zero, we must take $\zeta = 2$, i.e. $\lambda = 0$, to cancel the poles. Then
		\begin{gather}	
		g_{+-}(x) = x u_{+-}(1),\quad
		g_{-+}(x) = x^{L-1} u_{+-}(1).
		\end{gather}
	Clearly, this gives $u_{+-}(n) = \delta_{n,1} u_{+-}(1)$ and $u_{-+}(n) = \delta_{n,L-1} u_{+-}(1)$, which are the jammed steady states of the respective orientation sector. Obviously, for $\lambda = 0$ we will find $u_{++}(n) = u_{++}(1)$. For other eigenvalues, consistency requires $u_{+-}(1) = 0$ and the eigenvalues are associated exclusively with the dynamics of the $++$ (or $--$) orientation sector. These eigenvalues are just \eqref{eq:lambda_asym} with $\omega = 0$.

	\subsection{Spectrum: band structure and eigenvalue crossings}\label{sec:N2_spectrum}

	\begin{figure}[t]
		\begin{subfigure}{0.5\textwidth}
			\includegraphics[scale=0.8]{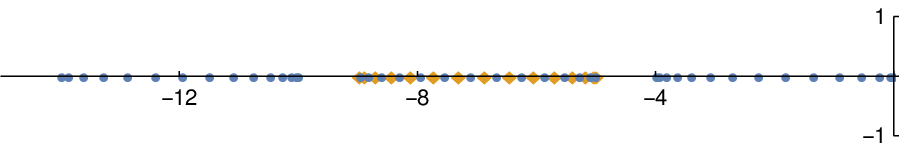}
			\caption{$\omega = 2.5$}
		\end{subfigure}
		\begin{subfigure}{0.5\textwidth}
			\includegraphics[scale=0.8]{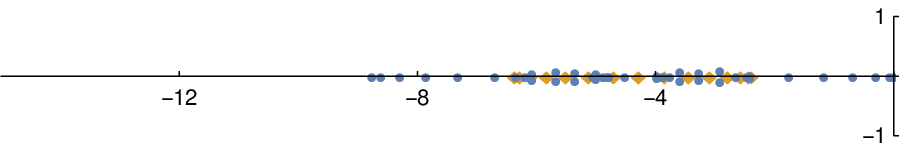}
			\caption{$\omega = 1.2$}\label{fig:N2_spec_w1p2}
		\end{subfigure}
		\begin{subfigure}{0.5\textwidth}
			\includegraphics[scale=0.8]{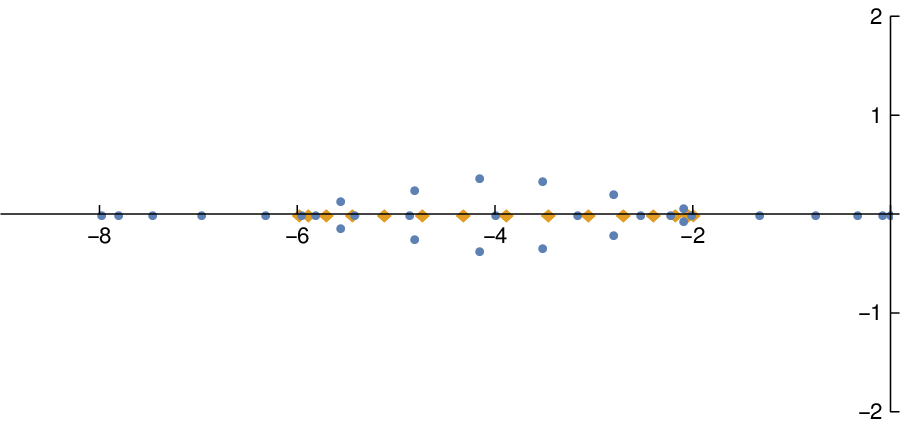}
			\caption{$\omega = 1$}
		\end{subfigure}
		\begin{subfigure}{0.5\textwidth}
			\includegraphics[scale=0.8]{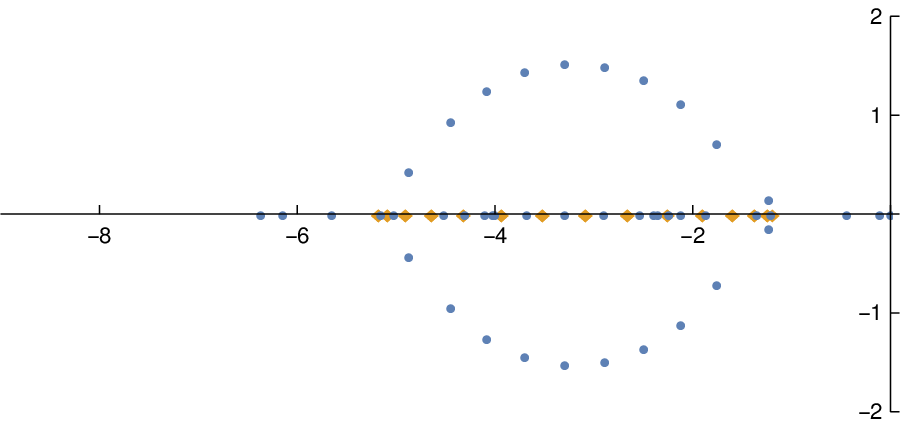}
			\caption{$\omega = 0.6$}
		\end{subfigure}		
		\begin{subfigure}{0.5\textwidth}
			\includegraphics[scale=0.8]{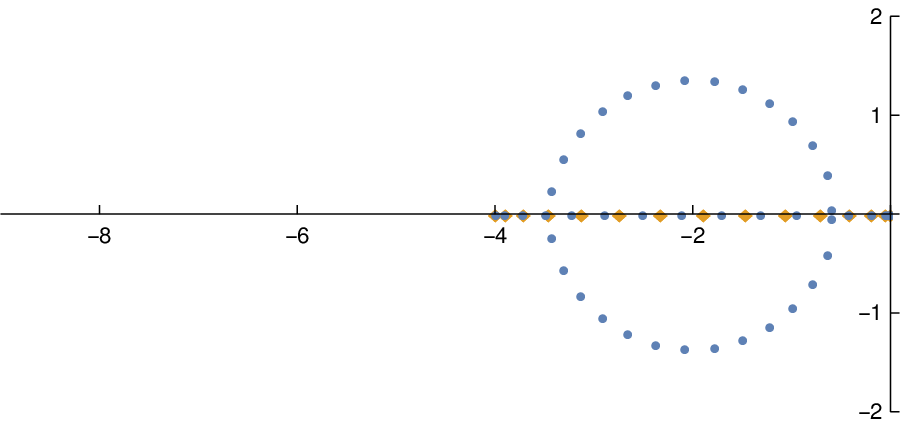}
			\caption{$\omega = 0.005$}
		\end{subfigure}
		\begin{subfigure}{0.5\textwidth}
			\includegraphics[scale=0.8]{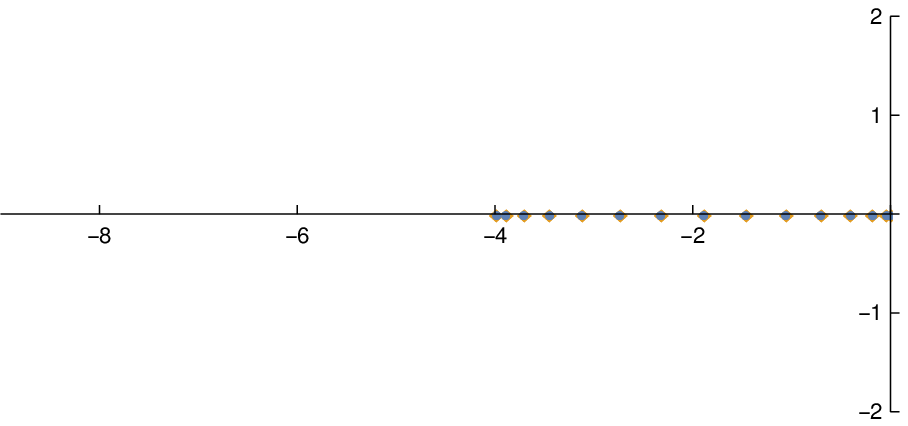}
			\caption{$\omega = 0$}
		\end{subfigure}		
		\caption{Plot of two-particle spectrum in the complex plane for $L=30$. Different $L$ produce the same pattern of eigenvalues. Blue circles: symmetric sector; orange diamonds: antisymmetric sector. (a) The symmetric sector comes in three separate bands for $\omega \geq 2$ where the separation is linear in $\omega$. (b) In the range $2\geq \omega > 1$ the bands cross, creating small excursions into the complex plane. (c) At exactly $\omega = 1$ there is degeneracy proportional to $L$ at $\lambda = -4$ as a `macroscopic eigenvalue crossing' takes place, just as in the one-particle case. (d) As the bands continue to cross, pairs of eigenvalues are sent out onto a deformed circle. (e) As $\omega$ approaches zero, the circle collapses towards $\lambda = -2$. (f) At $\omega = 0$ the symmetric and antisymmetric sector have the same eigenvalues (except the zero eigenvalue of the steady state which lies in the symmetric sector).}\label{fig:N2_spectrum}
	\end{figure}
	
	We now study the spectrum as a function of $\omega$ and $L$ by a combination of numerics and analytical results. For generic parameter values we find $2(L-1)$ eigenvalues, which is half the dimension of the Markov matrix. This is expected since we are restricted to the two $s=+1$ sectors out of four. The picture that emerges is plotted and described in detail in Figure \ref{fig:N2_spectrum}. For large $\omega$ the spectrum is structured into real bands which interact via eigenvalue crossings as $\omega$ tends to zero producing a complicated pattern.
	
	The spectrum is pieced together from the following results. The steady state and the two $L$-independent tumble modes with $\lambda = -2\omega, -4\omega$ were found analytically in the previous section, as well as the entire antisymmetric sector of eigenvalues \eqref{eq:lambda_asym} for arbitrary $L$ and $\omega > 0$. For $\omega = 0$ the formula \eqref{eq:lambda_asym} is again applicable but the spectrum is doubly degenerate (except the zero eigenvalue). For the symmetric sector, the eigenvalues are given by solving  \eqref{eq:polyeqs} and substituting their solution into \eqref{eq:lambda_of_z1}. In general, we do this numerically. While the roots generally fall on or near clear contours in the complex plane, we have omitted plotting the solutions to the roots in favour of plotting the resulting spectrum. For $\omega > 2$ it is possible to obtain an asymptotic analytic solution to the eigenvalue equations, derived in Appendix \ref{app:asymptotic}. This is facilitated by the numerical observation that in this region the roots form three groups, one with $z_1$ on the unit circle and $z_2$ real, and vice versa for the other two groups. Using this as an ansatz we find three real bands of eigenvalues
	\begin{subequations}
	\begin{alignat}{2}
	\lambda_m &= - 4 \sin^2 \theta_m - \frac{\sin^2 2\theta_m}{\omega} + \frac{2 \sin^2 2 \theta_m}{L\omega}  + \text{h.o.t.},&&\quad m\text{ integer},\\	
	\lambda_m &= - 4 \sin^2 \frac{\theta_m}{2} - 2\omega + \frac{ 2 \sin \theta_m \sin 2\theta_m}{L\omega^2}   + \text{h.o.t.},&& \quad m \text{ odd integer},\\
	\lambda_m &= - 4 \sin^2 \theta_m - 4\omega + \frac{\sin^2 2\theta_m}{\omega} - \frac{2 \sin^2 2 \theta_m}{L\omega} 
	 + \text{h.o.t.},&&\quad m\text{ integer},
	\end{alignat}
	\end{subequations}
	where $\theta_m = \pi m / (L-1)$ and `h.o.t' signifies terms with higher order reciprocals of $L$ or $\omega$.

	As a side remark: for $L \sim 30$, we have constructed the Markov matrix explicitly on a computer and solved numerically for the spectrum and eigenvectors. Those results are fully consistent with what we obtained by independent means, as presented in this section and the next. The naive numerical approach has the drawback of requiring all eigenvectors to be found and their symmetries determined, in order to select only the eigenvalues belonging to the two relevant symmetry sectors.

	\subsection{Right eigenvectors}\label{sec:N2_eigenvectors}
	
	The generating function expressed in the roots $z_1$ and $z_2$ is inverted in Appendix \ref{app:eigfun} to yield the eigenvectors. Here, we state the eigenvectors for any given eigenvalue $\lambda$, using \eqref{eq:def_z1z2} to define the roots from the shifted eigenvalue $\zeta = \lambda + 2(1+\omega)$. 
	
	Take first $\omega > 0$. The eigenvectors in the antisymmetric sector are given by
	\begin{subequations}\label{eq:antieigs}
		\begin{gather}
		u_{++}(n) = z_1^{n} + z_1^{L-n}  \\ 
		u_{+-}(n) = u_{-+}(n) = 0 \\		
		u_{--}(n) = - u_{++}(n).		
		\end{gather}
	\end{subequations}
	Since in this sector $z_1$ are $(L-1)$-roots of unity, the non-zero components are essentially Fourier basis functions. The symmetric sector has eigenvectors
	\begin{dgroup}\label{eq:solution}
		\begin{dmath}
		u_{++}(n) = \nu(z_1) \frac{ z_1 ^n + z_1^{L-n}}{1 - z_1^L} - \omega^2 \frac{ z_2 ^n + z_2^{L-n}}{\mu(z_2) + \nu(z_2)z_2^L},	
		\end{dmath}
		\begin{dmath}
		u_{+-}(n) = \omega \left(1 + \tfrac{2}{\zeta - 2}\delta_{n,1}\right) \left[ \frac{z_1^n - z_1^{L - n}}{1 - z_1^L} + \frac{ \nu(z_2) z_2^n + \mu(z_2) z_2^{L - n} }{\mu(z_2) + \nu(z_2)z_2^L} \right]
		\end{dmath}
		\begin{dmath}
		u_{-+}(n) = u_{+-}(L-n),	
		\end{dmath}
		\begin{dmath}
		u_{--}(n) = u_{++}(n),
		\end{dmath}
	\end{dgroup}
	where $\mu(z) = z - \zeta/2$ and $\nu(z) = \mu(1/z)$. The eigenvectors have two components, one involving $z_1$ and one $z_2$. In addition, the opposed orientation components have an `anomalous weight' on the jammed states $u_{+-}(1)$ and $u_{-+}(L-1)$. A few eigenvectors are shown in Figure \ref{fig:N2_eigenvectors}.

	\begin{figure}[t]
		\centering
		\begin{subfigure}{0.33\textwidth}		
			\includegraphics{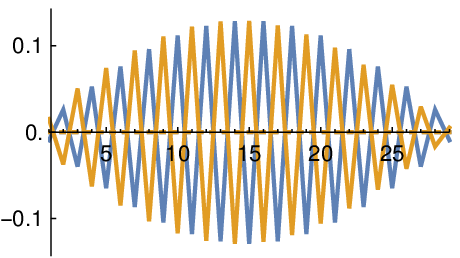}
			\caption{$\lambda = -8.78$}
		\end{subfigure}%
		\begin{subfigure}{0.33\textwidth}		
			\includegraphics{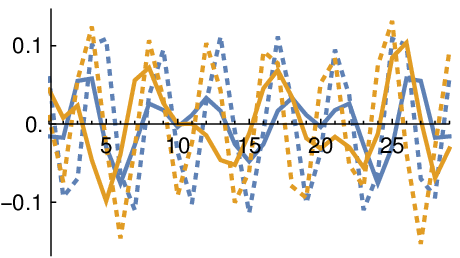}
			\caption{$\lambda = -2.92 - i0.0887$}
		\end{subfigure}%
		\begin{subfigure}{0.33\textwidth}		
			\includegraphics{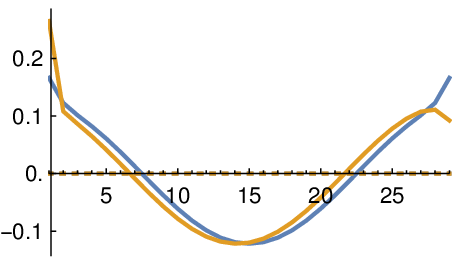}
			\caption{$\lambda = -0.0799$}
		\end{subfigure}
		\caption{Plot of some normalized eigenvectors for $L=30$, $\omega = 1.2$ (cf.\ Figure \ref{fig:N2_spec_w1p2}) versus lattice site $n$. Blue graphs: $u_{++}(n)$. Orange graphs: $u_{+-}(n)$. Full drawn lines show real part, dotted lines show imaginary part. (a) Most negative eigenvalue. (b) An arbitrary complex eigenvalue. (c) The least negative eigenvalue (spectral gap).}\label{fig:N2_eigenvectors}
	\end{figure}

	For $\omega = 0$, the orientation sectors are decoupled, so the relative scaling of the different orientation components is irrelevant,  
	\begin{subequations}
		\begin{align}
		u_{++}(n) &= z_1^{n} + z_1^{L-n}  \\ 
		u_{+-}(n) &\propto \delta_{\lambda,0} \delta_{n,1} \\
		u_{-+}(n) &\propto u_{+-}(L-n)  \\		
		u_{--}(n) &\propto u_{++}(n).		
		\end{align}
	\end{subequations}
	
	As a general observation on the basis of the uniqueness of the generating function inversion, a given eigenvalue cannot have distinct eigenvectors from the same sector. At an $\omega$ where two eigenvalues with eigenvectors in the same symmetry sector cross, the associated subspace becomes non-diagonalizable, just as in the one-particle case. However, we do not attempt to derive the generalized eigenvectors or projection operators for these cases. 
	
	\subsection{Nonequilibrium steady state}
	The steady state distribution, $P^*_{\sigma_1\sigma_2}(n)$, has been found and discussed in-depth by Slowman et al.\ \cite{Slowman2016a}, but for completeness we state it here in the context of the full spectral solution. 
	The steady state always lies in the symmetric sector (assuming $\omega > 0$ as we do throughout this section); being independent of initial condition, a symmetry argument implies $P_{++}^*(n) = P_{--}^*(n)$. As shown in Section \ref{sec:roots}, $z_1 = 1+\omega + \sqrt{\omega(2+\omega)}$ and $z_2 = 1$. The steady state distribution is then given by
	\begin{dgroup}\label{eq:uss}
		\begin{dmath}
		P^*_{++}(n) = \frac{1}{Z_L} \left( (1+\omega) \frac{ z_1 ^n + z_1^{L-n}}{1 - z_1^L} + \omega\right),	
		\end{dmath}
		\begin{dmath}\label{eq:Ppm}
		P^*_{+-}(n) = \frac{1}{Z_L}\left(\omega + \delta_{n,1}\right) \left( \frac{z_1^n - z_1^{L - n}}{1 - z_1^L} + 1 \right)
		\end{dmath}
		\begin{dmath}
		P^*_{-+}(n) = P^*_{+-}(L-n),	
		\end{dmath}
		\begin{dmath}
		P^*_{--}(n) = P^*_{++}(n),
		\end{dmath}
	\end{dgroup}
	where the normalization is
	\begin{equation}
	Z_L = \sum_{n=1}^{L-1} \sum\limits_{\sigma_1,\sigma_2 \\ \in\{+,-\}} u_{\sigma_1\sigma_2}(n) = 4 \left[ \omega(L-1) + (1+z_1) \frac{1- z_1^{L-1}}{1-z_1^L} \right].
	\end{equation}

	\subsection{Longest relaxation time and dynamical transition}\label{sec:N2_relax}
	
	\begin{figure}[t]
		\centering
		\begin{tikzpicture}
		\node[anchor=south west,inner sep=0] (image) at (0,0) 				{\includegraphics[scale=0.9]{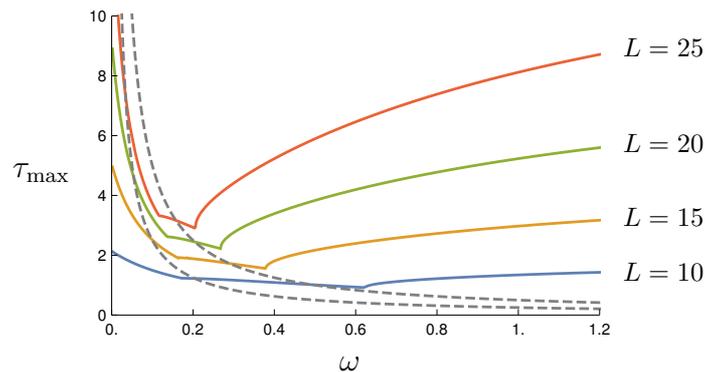}};
		\begin{scope}[x={(image.south east)},y={(image.north west)}]
		\node at (-0.09,0.5) {$\tau_{\text{max}}$};
		\node at (0.5,-0.08) {$\omega$};
		\node at (1.1, 0.88) {\footnotesize$L = 25$};
		\node at (1.1, 0.59) {\footnotesize$L = 20$};
		\node at (1.1, 0.365) {\footnotesize$L = 15$};
		\node at (1.1, 0.2) {\footnotesize$L = 10$};
		\end{scope}
		\end{tikzpicture}		
		\caption{ Coloured lines: longest system-size dependent relaxation time. Dashed lines: relaxation times $1/2\omega$ and $1/4\omega$.  Each coloured curve has two minima and cusps which signify dynamical transitions (cf.\ Figure \ref{fig:N1_tau_max}).}\label{fig:N2_tau_max}
	\end{figure}
		 
	The longest relaxation time $\tau_{\text{max}}$, disregarding the tumble modes, is obtained from the numerical solution of the spectrum as a function of $\omega$ and plotted in Figure \ref{fig:N2_tau_max}. While we do not have exact expressions for the values $\omega_{1,2}^*(L)$ at the two cusps, numerics suggests they scale as $\sim 1/L$ (for $L$ larger than about 20), in analogy with the one-particle case. The scaling away from the region between $\omega^*_1(L)$ and $\omega^*_2(L)$ is again $\sim L^2$. At present we do not have a physical picture to explain the emergence of a second cusp over the one-particle case.

	\section{Summary and discussion}\label{sec:sum}
	
	We have studied a model of interacting run-and-tumble particles on a ring lattice, and given the spectral solution for one and two particles. From a nonequilibrium theory point-of-view, this model is significant for exhibiting the following attributes: trivially breaking detailed balance (for $\omega < \infty$) due to persistent motion; breaking also kinematic reversibility due to the presence of jamming (for $N>1$); depending on a continuous-range parameter, the flipping rate $\omega$ (after choosing units of time such that $\gamma = 1$); and being to some extent analytically solvable as a function of $\omega$.  For both one and two particles we have found a rich spectral structure with intricate dependence on $\omega$ leading to qualitatively different relaxation regimes separated by an exceptional point, i.e.\ dynamical transitions. 
	
	The steady state for one particle is a uniform distribution, as is well known, and expected from the presence of kinematic reversibility and symmetry. Yet the non-stationary regime exhibits nonequilibrium effects since detailed balance is broken. For $\omega > 1$, the spectrum consists of two real bands that for smaller values combine in eigenvalue crossings to produce a complex-valued spectrum indicating oscillatory relaxation toward the steady state. Additionally, there is a tumble mode with $\lambda = -2\omega$, only involved in the orientation reversal dynamics. The dependence on $\omega$ of the longest relaxation time of the system (discounting the tumble mode) is characterized by a cusp minimum. The corresponding $\omega$ is an exceptional point where the dynamics is non-diagonalizable, and separates a region of regular from oscillatory exponential decay.
	
	For two particles, the lack of kinematic reversibility implies a non-trivial steady state \cite{Slowman2016a}. Our new spectral solution shows that, whilst being more intricate, the spectrum generalizes suggestively from the one-particle case. It has two sectors, corresponding to the allowed symmetries of the eigenvectors. The larger sector consists for $\omega > 2$ of three separate real bands, which for smaller $\omega$ engage in eigenvalue crossings to become complex, in general. Except at eigenvalue crossings, where once again the dynamics is non-diagonalizable, the spectrum is determined by the two polynomial equations in \eqref{eq:polyeqs}, expressed in the roots $z_1$ and $z_2$ \eqref{eq:z1def}, which are functions of the eigenvalue. The other sector is a shift by $-2\omega$ of the TASEP spectrum (excluding just the 0 eigenvalue). There are now two tumble modes, $\lambda = -4\omega, -2\omega$, one in each symmetry sector. (It is to be expected, but was not  actually proven, that they are uninvolved in the spatial dynamics.) The longest relaxation time (excluding tumble modes) has cusp minima at two exceptional values of $\omega$. The eigenvectors of the larger symmetry sector, given by \eqref{eq:solution}, have a Bethe-esque structure in terms of the roots, and are quasi-sinusoidal but with an anomalous weight put on the jammed configurations. 	

	The general $N$-particle problem remains unsolved. At present, we do not know if the generating function technique used in this article, involving only one spatial coordinate, generalizes in a tractable way to more coordinates. Still, we may guess at some features of the general solution. Certainly, the spectrum will be divided into some number of symmetry sectors. We would expect to find branch-point type eigenvalue crossings; these have been proven to exist in the $N$-particle Bethe ansatz solution of the ASEP \cite{Brattain2016} (there referred to as `ramification points'). Generalizing the obtained results rather naively, we conjecture that the spectrum consists of a number real bands for $\omega \gtrsim N$, which cross in complicated ways for smaller $\omega$. Tumble modes related only to the orientation dynamics may still occur, proposedly at $\lambda = -2k$, for $k = 1,2,\ldots,N$. In the solved cases, the dynamical transitions occurred at special values $\omega^*\sim 1/L$. These would disappear if $L$ was taken to infinity, and are in that sense a finite-size effect. However, the more relevant limit is to keep $N/L$ finite, and the dynamical transitions may persist. Note, however, that such a transition would not be the effect of the said limit, since the non-analyticity of the relaxation time itself exists for finite particle number.

	In conclusion, we have shown that even a simple nonequilibrium model, such as the single run-and-tumble particle, can have an intricate spectral structure as a function of a real parameter. The further richness of the two-particle solution highlights the difficulty of solving interacting random walk models, but does not exclude the possibility of a general $N$-particle solution. We expect exceptional points of the parameter space to be ubiquitous in nonequilibrium models, and possibly prove to be of similar importance as has been realized in other areas of physics over the past 15 years \cite{Heiss2012}.
	
	\section*{Acknowledgements}\label{sec:ack} \addcontentsline{toc}{section}{Acknowledgements}
	Emil Mallmin acknowledges studentship funding from EPSRC grant no.\ EP/N509644/1.

	\begin{appendices}
	
	\section{Asymptotic solution of two-particle spectrum for $\omega > 2$}\label{app:asymptotic}
	
	Here, we solve the polynomial equations \eqref{eq:polyeqs} to order $1/L$, in the regime $\omega > 2$. Guided by the numerical solution, we make the ansatz that either $z_1$ or $z_2$ lies on the unit circle. The first assumption produces the middle band, and the second ansatz the left and right bands.   
	
	\subsection{The middle band}
	
	Starting with an ansatz for $z_1$ on the unit circle, we assume it differs from a root of negative one by an argument of order $1/L$,
	\begin{equation}\label{eq:z1exp}
	z_1 = \exp \left[ i\left( \theta - \frac{f(\omega,\theta)}{L} + \Ordo{L^{-2}}\right) \right],
	\end{equation}
	where $\theta = \theta_m = \pi m /(L-1)$ and $m$ is an odd integer up to $2(L-1)-1$. We rewrite \eqref{eq:eigeneq1} as		
	\begin{equation}\label{eq:zpoly1_rho}
	z_2^{L-1}  = \frac{\mu(z_2)\mu(z_1) + \omega^2 \rho(z_1)}{\nu(z_2)\mu(z_1) + \omega^2 \rho(z_1)},\quad \rho(z_1) = \frac{1 + z_1^{L-1}}{1 - z_1^{L-1}},
	\end{equation}	
	which stands in for \eqref{eq:zpoly1}. To leading order, $z_1^{L-1} = - e^{-i f}$, so that
	\begin{equation}\label{eq:rhotan}
	\rho(z_1) = \frac{1 - e^{-i f(\omega,\theta)}}{1 + e^{-if(\omega,\theta)}} = i T(\omega, \theta),\quad T(\omega,\theta) = \tan [f(\omega,\theta)/2].
	\end{equation}
	Using the fact that $z_1$ is on the unit circle, we rewrite \eqref{eq:zpoly2}  as
	\begin{equation}\label{eq:poly2z1Re}
	\bar{z}_2 = - z_2 + \frac{1-\omega^2}{\Re z_1} + \Re z_1.
	\end{equation}
	Note that $\mu(x) = x - \Re z_1 $, so that $\mu(z_1) = i \Im z_1$. Then substituting \eqref{eq:poly2z1Re} and \eqref{eq:rhotan} into \eqref{eq:zpoly1_rho}, one obtains
	\begin{equation}\label{eq:z2alteq}
	z_2^{L-1} = - \frac{z_2 - \Re z_1 + \omega^2 T/\Im z_1}{z_2 - 1/\Re z_1 - \omega^2 \left(T/\Im z_1 - 1/\Re z_1\right)}.
	\end{equation} 	
	Due to the reciprocal symmetry, we can without loss of generality take $|z_2|{} < 1$ . Then $\lim\limits_{L\to \infty} z_2^{L-1} = 0$, which implies that the numerator in the r.h.s.\ above vanishes in this limit. Hence
	\begin{equation}\label{eq:z2hat}
	\hat{z}_2 \equiv \liminfty{L} z_2 = \cos \theta - \omega^2 \frac{T(\omega,\theta)}{\sin \theta}.
	\end{equation}
	We take now the $L\to\infty$ limit of \eqref{eq:poly2z1Re}, and substitute the expression \eqref{eq:z2hat} for $\hat{z}_2$. Introducing the small parameter $\varepsilon = 1/\omega^2$,
	\begin{equation}
	\left[\cos \theta - \frac{T}{\varepsilon \sin \theta}\right]^{-1} - \frac{T}{\varepsilon \sin \theta} = \frac{\varepsilon - 1}{\varepsilon \cos \theta}.
	\end{equation}
	Straightforward manipulations lead to the quadratic equation
	\begin{equation}
	T^2 - [\varepsilon \sin \theta \cos \theta + (1 - \varepsilon)\tan \theta] T + \varepsilon \sin^2 \theta = 0,
	\end{equation}
	with solution
	\begin{equation}
	T(\omega,\theta) = \frac{1}{2}\left[ \varepsilon \sin \theta \cos\theta + (1-\varepsilon) \tan \theta \right] \pm \sqrt{\frac{1}{4}\left[ \varepsilon \sin \theta \cos\theta + (1-\varepsilon) \tan \theta \right]^2 - \varepsilon \sin^2 \theta}.
	\end{equation}
	The choice of root relates to the reciprocal symmetry of the solutions, and we can without loss of generality take the consistent combination of negative root and $\tan \theta > 0$. 
	
	Although $f(\omega,\theta)$ has now been found exactly, the expression is unwieldy. Settling for truncated Laurent series in $\omega$, a computer algebra package finds for us
	\begin{equation}\label{eq:f}
	f(\omega,\theta) = \frac{1}{\omega^2}\left(1 + \frac{1}{\omega^2}\right) \sin 2\theta + \Ordo{\omega^{-6}}.
	\end{equation}	
	The corresponding eigenvalue is given by $\lambda = 2 \Re z_1 - 2(1+\omega)$. Substituting in \eqref{eq:z1exp} and \eqref{eq:f}, and expanding up to relevant orders,
	\begin{equation}
	\lambda_m = - 4 \sin^2 \frac{\theta_m}{2} - 2\omega + \frac{ 2 \sin \theta_m \sin 2\theta_m}{L\omega^2}   + \text{h.o.t.},
	\end{equation}
	where $\text{h.o.t.}$ implies terms with higher reciprocal orders of $\omega$ or $L$.
	
	\subsection{The left and right bands}
	
	For values of $z_1$ that lie on the real axis, the corresponding $z_2$ is instead on the unit circle. In particular, there are are two distinct $z_2$'s (with different $z_1$'s) close to every root of positive one. We make the ansatz
	\begin{equation}
	z_2 = \exp\left[i\left( \varphi_m - \frac{h(\omega,\varphi_m)}{L} + \Ordo{1/L^2} \right)\right],	
	\end{equation}	
	where $\varphi = \varphi_m = 2\pi m /(L-1)$, $m = 1,2,\ldots,L-1$. We can then proceed in much the same way as with the previous ansatz, but interchanging the role of $z_1$ and $z_2$. First, \eqref{eq:zpoly1} is written
	\begin{equation}\label{eq:z1polyEta}
	z_1^{L-1} = - \frac{\omega^2 + \mu(z_1) \eta(z_2)}{\omega^2 + \nu(z_1) \eta(z_2)},\quad \eta(z_2) = \frac{\mu(z_2) - \nu(z_2)z_2^{L-1}}{1 - z_2^L},
	\end{equation}
	and \eqref{eq:zpoly2} is solved
	\begin{equation}\label{eq:qz2UC}
	z_1 + \bar{z}_1 = 2 \left[ \Re z_2 + q\sqrt{\omega^2 - (\Im z_2)^2} \right],
	\end{equation}
	where $q = \pm 1$ selects between the two solutions. Then, since $\zeta = z_1 + \bar{z}_1$, we have 
	\begin{gather}
	\mu(z_2) = i \Im z_2 - q \sqrt{\omega^2 - (\Im z_2)^2}, \\
	\nu(z_2) = -i \Im z_2 - q \sqrt{\omega^2 - (\Im z_2)^2}.
	\end{gather}	
	It follows that (to leading order)
	\begin{equation}\label{eq:etaz2}
	\begin{split}
	\eta(z_2) &= \frac{i (\Im z_2) (1 + z_2^{L-1}) - q \sqrt{\omega^2 - (\Im z_2)^2} (1- z_2^{L-1})}{1- z_2^{L-1}} \\
	&= i (\Im z_2) \frac{1 + z_2^{L-1}}{1 - z_2^{L-1}} - q \sqrt{\omega^2 - (\Im z_2)^2} \\
	&= \frac{\Im z_2}{T(\omega,\varphi)} - q \sqrt{\omega^2 - (\Im z_2)^2},
	\end{split}
	\end{equation}
	where $T(\omega,\varphi)=\tan [h(\omega,\varphi)/2]$. Without loss of generality we assume $|z_1|{} < 1$, so that taking the limit $L\to \infty$ of \eqref{eq:z1polyEta} leads to
	\begin{equation}
	\mu(\hat{z}_1) = - \frac{\omega^2}{\eta(\hat{z}_2)}.
	\end{equation} 
	This we express using \eqref{eq:qz2UC} and \eqref{eq:etaz2} as
	\begin{equation}
	\hat{z}_1 = \cos \varphi + q \sqrt{\omega^2 - \sin^2 \varphi} - \frac{\omega^2 T(\omega,\varphi)}{\sin \varphi - q \sqrt{\omega^2 - \sin^2 \varphi} T(\omega,\varphi)}.
	\end{equation}
	
	Now, going back to \eqref{eq:qz2UC}, we solve for $\hat{z}_1$ to obtain
	\begin{equation}
	\hat{z}_1 = \cos \varphi + q \sqrt{\omega^2 - \sin^2 \varphi} + q' \sqrt{\left(  \cos \varphi + q \sqrt{\omega^2 - \sin^2 \varphi} \right)^2 - 1},
	\end{equation}
	where $q' = \pm 1$ selects the positive or negative root solution. Dividing by $\omega$ and taking the limit to infinity, we discover the consistency requirement $q' = -q$. Combining the last two equations,
	\begin{equation}
	\frac{\omega^2 T(\omega,\varphi)}{\sin \varphi - q \sqrt{\omega^2 - \sin^2 \varphi} T(\omega,\varphi)} = q \sqrt{\left(\cos \varphi + \sqrt{\omega^2 - \sin^2 \varphi} \right)^2 - 1}.
	\end{equation}	
	After rearranging we find
	\begin{equation}
	h(\omega,\varphi) = 2 \arctan \left[ \frac{q \sin \varphi}{\omega^2 \left[\left(\cos \varphi + q \sqrt{\omega^2 - \sin^2 \varphi}\right)^2 - 1\right]^{-1/2} + \sqrt{\omega^2 - \sin^2 \varphi} }  \right].
	\end{equation}
	This we expand using computer algebra to
	\begin{equation}
	h(\omega, \varphi) =  \frac {q \sin \varphi}{\omega}+\frac{\sin \varphi \cos
		\varphi}{2 \omega^2} -\frac{ q  \sin \varphi\, (\cos 2 \varphi
		+5)}{12 \omega^3}+\Ordo{\omega^{-4}}.
	\end{equation}
	Once $z_2$ is known, the eigenvalue is obtained from \eqref{eq:qz2UC} and expanded,
	\begin{equation}\label{eq:lambda_s_bands}
	\lambda_m = - 4 \sin^2 \frac{\varphi}{2} - 2\omega(1-q)- \frac{q \sin^2 \varphi}{\omega} + \frac{2q \sin^2 \varphi}{L\omega} + \frac{3 \sin \varphi \sin 2 \varphi}{2 L \omega^2} + \text{h.o.t.}
	\end{equation}

	\section{Derivation of two-particle eigenvectors}\label{app:eigfun}
	
	Here we determine the generating function by evaluating the inversion \eqref{eq:g_H_inv}. We begin with the symmetric sector. Define for convenience
	\begin{equation}\label{eq:N2_J_pole}
	J(x) = \mu(x)\nu(x)u_{++}(1) - \omega \mu(x) u_{+-}(1).
	\end{equation}
	The first component of the generating function can then be expressed
	\begin{equation}
	g_{++}(x) = - \frac{2 x^2(1-x)}{\zeta(x - z_1)(x - \tfrac{1}{z_1})(x - z_2)(x - \tfrac{1}{z_2})}[J(x) - x^{L-1}J(1/x)].
	\end{equation}
	Since $J(1/x) \sim 1/x$, it follows that $x^2(1-x)J(1/x) \sim x$, and therefore the second term is $\sim x^L$ once the factors in the denominator are expanded in geometric series. It is thus unimportant, since by the definition of $g(x)$ all powers above $L-1$ will eventually cancel out.  Hence we write
	\begin{equation}\label{eq:gpp}
	g_{++}(x) = - \frac{2x}{\zeta} \cdot \frac{x(1-x)J(x)}{(x - z_1)(x - \tfrac{1}{z_1})(x - z_2)(x - \tfrac{1}{z_2})}+ \Ordo{x^L}.
	\end{equation}
	We perform a partial fraction decomposition of the large fraction, possible since the numerator is $\sim x^3$;
	\begin{dmath}
	g_{++}(x) = - \frac{2x}{\zeta} \left\{ \frac{z_1(1-z_1)J(z_1)}{(x - z_1)(z_1 - \tfrac{1}{z_1})(z_1 - z_2)(z_1 - \tfrac{1}{z_2})} \\+ \frac{\sfrac{1}{z_1}(1-\sfrac{1}{z_1})J(\sfrac{1}{z_1})}{(\tfrac{1}{z_1} - z_1)(x - \tfrac{1}{z_1})(\tfrac{1}{z_1} - z_2)(\tfrac{1}{z_1} - \tfrac{1}{z_2})} + \text{perm.}\right\} + \Ordo{x^L},
	\end{dmath}
	where `perm.' implies a repetition of the terms to its left but with $z_1$ and $z_2$ permuted. The pole cancellation conditions \eqref{eq:Cb_zero} imply succinctly 
	\begin{equation}\label{eq:Jpolecond}
	J(z_i) = z_i^{L-1} J(1/z_i),\quad i = 1, 2.
	\end{equation}
	Using \eqref{eq:Jpolecond}, together with the algebraic identity
	\begin{equation}
	(\tfrac{1}{z_1} - z_2)(\tfrac{1}{z_1} - \tfrac{1}{z_2}) = \tfrac{1}{z_1^2}(z_1 - z_2)(z_1 - \tfrac{1}{z_2}),
	\end{equation}
	the expression \eqref{eq:gpp} simplifies to
	\begin{equation}
	g_{++}(x) = \frac{2 (1-z_1)J(\sfrac{1}{z_1})}{\zeta(z_1 - \tfrac{1}{z_1})(z_1 - z_2)(z_1 - \tfrac{1}{z_2})}  \left[ \frac{z_1^L x}{z_1 - x} + \frac{x}{\tfrac{1}{z_1} - x} \right]+ \text{perm.} + \Ordo{x^L}.
	\end{equation}
	Recognizing the power series
	\begin{equation}\label{eq:geom}
	\frac{x}{a - x} = \sum_{n = 1}^\infty (x/a)^n,
	\end{equation}
	we have found
	\begin{equation}
	g_{++}(x) = \frac{2(1-z_1)J(\sfrac{1}{z_1})}{\zeta(z_1 - \tfrac{1}{z_1})(z_1 - z_2)(z_1 - \tfrac{1}{z_2})}   \sum_{n=1}^{L-1} [z_1^n + z_1^{L-n}] x^n 
	+ \text{perm.},
	\end{equation}
	since higher order terms must cancel out. Denote the prefactors by
	\begin{equation}\label{eq:A1A2}
	A_1 \nu(z_1) = \frac{2(1-z_1)J(\sfrac{1}{z_1})}{\zeta(z_1 - \tfrac{1}{z_1})(z_1 - z_2)(z_1 - \tfrac{1}{z_2})} ,\quad -A_2 \omega = \frac{2(1-z_2)J(\sfrac{1}{z_2})}{\zeta(z_2 - \tfrac{1}{z_2})(z_2 - z_1)(z_2 - \tfrac{1}{z_1})}.
	\end{equation}
	Then
	\begin{equation}\label{eq:uppprel}
	u_{++}(n) = A_1 \nu(z_1)(z_1^n + z_1^{L-n} ) - A_2 \omega (z_2^n + z_2^{L-n}).
	\end{equation}
	
	To find the second component, $u_{+-}(n)$, we define for convenience the two functions
	\begin{gather}
	K(x) = \omega \mu(x) u_{++}(1) + \omega^2 u_{+-}(1),\\
	\begin{split}\label{eq:Kprime}
	K'(x) &= \mu(x)(\mu(x) + \nu(x) )u_{+-}(1) \\
	&= \frac{ \mu(x)}{x}(x - z_1)(x - \sfrac{1}{z_1}) u_{+-}(1).
	\end{split}
	\end{gather}
	With these definitions
	\begin{equation}
	g_{+-}(x) = \frac{2x}{\zeta}\cdot \frac{x(1-x) }{(x-z_1)(x-\tfrac{1}{z_1})(x-z_2)(x-\tfrac{1}{z_2})}[K(x) - K'(x)] + \Ordo{x^L}.
	\end{equation}	
	The $K(x)$ term can be decomposed by four partial fractions since $K(x) \sim x$, leaving the numerator $\sim x^3$. The $K'(x)$ term already cancels two of the poles and $x$ in the numerator, after which two partial fractions can be taken. The result is
	\begin{dmath}
	g_{+-}(x) = - \frac{2 (1-z_1)}{\zeta(z_1 - \tfrac{1}{z_1})(z_1 - z_2)(z_1 - \tfrac{1}{z_2})} \left[ \frac{K(\sfrac{1}{z_1}) x }{\sfrac{1}{z_1} - x} + \frac{K(z_1)z_1 x}{z_1 - x} \right] + \text{perm.} \\ +   \frac{2(1-x)u_{+-}(1)}{\zeta (z_2 - \tfrac{1}{z_2})} \left[ \mu(z_2)\frac{x}{z_2 - x} - \nu(z_2)\frac{x}{\sfrac{1}{z_2} - x} \right]
	\end{dmath}	
	We make use of the following relations,
	\begin{gather}
	K(z_1) = \frac{\omega}{\nu(z_1)} z_1^{L-1}J(1/z_1),\\
	K(1/z_1) = - \frac{\omega}{\nu(z_1)}J(1/z_1),\\
	K(z_2) =  \frac{\mu(z_2)}{\omega} z_2^{L-1}J(z_2) + K'(z_2),\\
	K(\sfrac{1}{z_2}) = \frac{\nu(z_2)}{\omega} J(\sfrac{1}{z_2}) + \frac{\nu(z_2)}{\mu(z_2)} K'(z_2),
	\end{gather}
	and the geometric series \eqref{eq:geom}, to find	
	\begin{dmath}
	g_{+-}(x) = A_1 \omega \sum_{n=1}^{\infty}[z_1^n - z_1^{L-n}]x^n +  A_2 \sum_{n=1}^\infty [\nu(z_2) z_2^n + \mu(z_2)z_2^{L-n}]x^n  - \frac{2 (1 - z_2) K'(z_2)}{\zeta \mu(z_2) (z_2 - \tfrac{1}{z_2})(z_2 - z_1)(z_2 - \tfrac{1}{z_1})} \left[ \nu(z_2) \frac{x}{\sfrac{1}{z_2} - x} + \mu(z_2
	)\frac{z_2 x}{z_2 - x}  \right] +   \frac{2(1-x)u_{+-}(1)}{\zeta (z_2 - \tfrac{1}{z_2})} \left[ \mu(z_2)\frac{x}{z_2 - x} - \nu(z_2)\frac{x}{\sfrac{1}{z_2} - x} \right].
	\end{dmath}		
	Considering \eqref{eq:Kprime}, the last two lines almost cancel, leaving only a term $(2/\zeta) x u_{+-}(1)$. We have then found	
	\begin{equation}\label{eq:upmprel}
	u_{+-}(n) =  A_1 \omega (z_1^n - z_1^{L-n}) + A_2 (\nu(z_2) z_2^n + \mu(z_2)z_2^{L-n}) + \frac{2}{\zeta} u_{+-}(1) \delta_{n,1}.
	\end{equation}
	This we can rearrange to
	\begin{equation}\label{eq:upm_app}
		u_{+-}(n) = \left(1+ \frac{2}{\zeta - 2} \delta_{n,1} \right) \left( A_1 \omega (z_1^n - z_1^{L-n}) + A_2 (\nu(z_2) z_2^n + \mu(z_2)z_2^{L-n}) \right) .
	\end{equation}
	
	Finally, we want to choose the arbitrary overall scaling of the eigenvectors such that $A_1$ and $A_2$ become simple expressions. The ratio $A_1/A_2$ is already fixed by $u_{++}(1)/u_{+-}(1)$ which must satisfy \eqref{eq:eigeneq1}. Rather than attempting to simplify \eqref{eq:A1A2} directly, we substitute \eqref{eq:uppprel} and \eqref{eq:upmprel} into the eigenvalue equation corresponding to \eqref{eq:N2_ME_-+} for $n=1$, which reads simply	
	\begin{equation}
	\frac{\zeta}{2} u_{-+}(1) = \omega u_{++}(1).
	\end{equation}
	After trivial rearrangements, the above equation gives
	\begin{equation}\label{eq:A1A2_eq}
	A_1 \left[ \frac{\zeta}{2}(z_1 - z_1^{L-1}) + \nu(z_1)(z_1 + z_1^{L-1}) \right] = A_2 \frac{1}{\omega} \left[ \frac{\zeta}{2}( \mu(z_2) + \nu(z_2) z_2^{L-1}) + \omega^2 (z_2 + z_2^{L-1}) \right].
	\end{equation}
	The bracket on the left-hand side can be written using $\nu(z_1) = - \mu(z_1)$ \eqref{eq:N2_z_roots_def} as
	\begin{equation}
	\left( \frac{\zeta}{2}z_1 + \nu(z_1)z_1 \right) - \left( \mu(z_1) + \frac{\zeta}{2}\right) z_1^{L-1}.
	\end{equation}
	After using the definitions of $\mu$ and $\nu$ \eqref{eq:mu_nu}, the resulting expression is
	\begin{equation}
	1 - z_1^L.
	\end{equation}
	For the left-hand-side bracket in \eqref{eq:A1A2_eq} we use $\omega^2 = \mu(z_2)\nu(z_2)$ \eqref{eq:N2_z_roots_def} to express it as
	\begin{equation}
	\mu(z_2)\left( \frac{\zeta}{2}z_2 + \nu(z_2)z_2\right) + \nu(z_2) \left( \frac{\zeta}{2} + \mu(z_2) \right)z_2^{L-1}.
	\end{equation}
	After simplifying the above using \eqref{eq:mu_nu}, it is clear that we can choose
	\begin{equation}
	A_1 = \frac{1}{1 - z_1^L},\quad A_2 = \frac{\omega}{\mu(z_2) + \nu(z_2)z_2^L}.
	\end{equation}
	This concludes the derivation of eigenvectors for the symmetric sector.

	The non-zero eigenvector \eqref{eq:g++asym} of the asymmetric sector is obtained by the same method of partial fraction decomposition and geometric series expansion as above.
	
\end{appendices}
	
	\addcontentsline{toc}{section}{References}
	\bibliographystyle{physicsbibstyle.bst}	
		
	\bibliography{references_rtp_article.bib}     

\begin{thebibliography}{10}
\providecommand{\url}[1]{\texttt{#1}}
\providecommand{\urlprefix}{URL }

\bibitem{Einstein1905}
A.~Einstein.
\newblock \"Uber die von der molekular-kinetischen {Theorie} der {W\"arme}
  gefordete {Bewegung} von in ruhenden {Fl\"ussigkeiten} suspendierten
  {Teilchen}.
\newblock \emph{Ann. Phys.} \textbf{17}, 549 (1905)

\bibitem{Smoluchowski1906}
M.~von Smoluchowski.
\newblock Zeu kinetischen {Theorie} der {Brownsche} {Bewegung}.
\newblock \emph{Ann. Phys.} \textbf{21}, 756 (1906)

\bibitem{Crooks2000}
G.~E. Crooks.
\newblock Parth-ensemble averages in systems driven far from equilibrium.
\newblock \emph{Phys. Rev. E} \textbf{61}, 3 (2000)

\bibitem{Klages2013}
R.~Klages, W.~Just, C.~Jarzynski (Editors) \emph{Nonequilibrium Statistical
  Physics of Small Systems}.
\newblock Wiley-VCH (2013)

\bibitem{Bechinger2016}
C.~Bechinger, R.~D. Leonardo, H.~L\"owen, C.~Reichhardt, G.~Volpe, G.~Volpe.
\newblock Active particles in complex and crowded environments.
\newblock \emph{Rev. Mod. Phys.} \textbf{88}, 046006 (2016)

\bibitem{Fodor2018}
E.~Fodor, C.~Marchetti.
\newblock The statistical physics of active matter: From self-catalytic
  colloids to living cells.
\newblock \emph{Physica A} \textbf{504}, 106 (2018)

\bibitem{Steffenoni2016}
S.~Steffenoni, K.~troy, G.~Falasco.
\newblock Interacting {Brownian} dynamics in a nonequilibrium particle bath.
\newblock \emph{Phys. Rev. E} \textbf{94}, 062139 (2016)

\bibitem{Cates2015}
M.~E. Cates, J.~Tailleur.
\newblock Motility-Induced Phase Separation.
\newblock \emph{Annu. Rev. condens. Matter Phys.} \textbf{6}, 219 (2015)

\bibitem{Bricard2013}
A.~Bricard, J.-B. Caussin, N.~Desreumaux, O.~Dauchot, D.~Bartolo.
\newblock Emergence of macroscopic directed motion in populations of motile
  colloids.
\newblock \emph{Nature} \textbf{503}, 95 (2013)

\bibitem{Hinrichsen2000}
H.~Hinrichsen.
\newblock Non-equilibrium critical phenomena and phase transitions into
  absorbing states.
\newblock \emph{Advances in Physics} \textbf{49}, 7 (2000)

\bibitem{Blythe2003}
R.~A. Blythe, M.~R. Evans.
\newblock The Lee-Yang Theory of Equilibrium and Nonequilibrium Phase
  Transitions.
\newblock \emph{Brazilian Journal of Physics} \textbf{33}, 3 (2003)

\bibitem{Evans2002}
M.~R. Evans, R.~Blythe.
\newblock Nonequilibrium dynamics in low-dimensional systems.
\newblock \emph{Physica A} \textbf{313}, 110 (2002)

\bibitem{Schmittmann1995}
B.~Schmittmann, R.~K.~P. Zia.
\newblock Statistical mechanics of driven diffusive systems.
\newblock In C.~Domb, J.~L. Lebowitz (Editors) \emph{Phase Transitions and
  critical Phenomena}, volume~17. San Diego: Academic (1995)

\bibitem{Derrida1998}
B.~Derrida.
\newblock An exactly soluble non-equilibrium system: The asymmetric simple
  exclusion process.
\newblock \emph{Phys. Rep.} \textbf{301}, 65 (1998)

\bibitem{Golinelli2006}
O.~Golinelli, K.~Mallick.
\newblock The asymmetric simple exclusion process: an integrable model for
  non-equilibrium statistical mechanics.
\newblock \emph{J. Phys. A: Math. Gen.} \textbf{39}, 12679 (2006)

\bibitem{Chou2011}
T.~Chou, K.~Mallick, R.~K.~P. Zia.
\newblock Non-equilibrium statistical mechanics: from a paradigmatic model to
  biological transport.
\newblock \emph{Rep. Prog. Phys.} \textbf{74}, 116601 (2011)

\bibitem{Berg2004}
H.~C. Berg.
\newblock \emph{E. coli in Motion}.
\newblock Springer (2004)

\bibitem{Masoliver1989}
J.~Masoliver, K.~Lindenberg, G.~H. Weiss.
\newblock A continuous-time generalization of the persistent random walk.
\newblock \emph{Physica A} \textbf{157}, 891 (1989)

\bibitem{Masoliver2017}
J.~Masoliver, K.~Lindenberg.
\newblock Continuous time persistent random walk: a review and some
  generalizations.
\newblock \emph{Eur. Phys. J. B} \textbf{90}, 107 (2017)

\bibitem{Weiss1994}
G.~H. Weiss.
\newblock \emph{Aspects and applications of the random walk}.
\newblock North-Holland (1994)

\bibitem{Boguna1999}
M.~{Bogu\~n\'a}, J.~M. Porr\`a, J.~Masoliver.
\newblock Persistent random walk model for transport through thin slabs.
\newblock \emph{Phys. Rev. E} \textbf{59}, 6 (1999)

\bibitem{Miri2003}
M.~Miri, H.~Stark.
\newblock Persistent random walk in a honeycomb structure: light transport in
  foams.
\newblock \emph{Phys. Rev. E} \textbf{68}, 031102 (2003)

\bibitem{Wu2000}
H.~Wu, B.-L. Li, T.~A. Springer, W.~H. Neill.
\newblock Modelling animal movement as a persistent random walk in two
  dimensions.
\newblock \emph{Ecological Modelling} \textbf{132}, 115 (2000)

\bibitem{Fujita1980}
S.~Fujita, Y.~Okamura.
\newblock Theory of polymer conformation based on the correlated walk model.
\newblock \emph{J. Chem. Phys.} \textbf{72}, 3993 (1980)

\bibitem{Bhat2013}
D.~Bhat, M.~Gopalakrishnan.
\newblock Memory, bias, and correlations in bidirectional transport of
  molecular-motor-driven cargoes.
\newblock \emph{Phys. Rev. E} \textbf{88}, 042702 (2013)

\bibitem{Schnitzer1993}
M.~J. Schnitzer.
\newblock Theory of continuum random walks and application to chemotaxis.
\newblock \emph{Phys. Rev. E} \textbf{48}, 4 (1993)

\bibitem{Angelani2017}
L.~Angelani.
\newblock Confined run-and-tumble swimmers in one dimension.
\newblock \emph{J. Phys. A: Math. Theor,} \textbf{50}, 325601 (2017)

\bibitem{Thompson2011}
A.~G. Thompson, J.~Tailleur, M.~E. Cates, R.~A. Blythe.
\newblock Lattice models of nonequilibrium bacterial dynamics.
\newblock \emph{J. Stat. Mech.} {P02029} (2011)

\bibitem{Tailleur2008}
J.~Tailleur, M.~E. Cates.
\newblock Statistical Mechanics of Interacting Run-and-Tumble Bacteria.
\newblock \emph{Phys. Rev. Lett.} \textbf{100}, 218103 (2008)

\bibitem{Tailleur2009}
J.~Tailleur, M.~E. Cates.
\newblock Sedimentation, trapping, and rectification of dilute bacteria.
\newblock \emph{EPL} \textbf{86}, 60002 (2009)

\bibitem{Escaff2018}
D.~Escaff, R.~Toral, C.~V. den Broeck, K.~Lindenberg.
\newblock A continuous-time persistent random walk model for flocking.
\newblock \emph{Chaos} \textbf{28}, 075507 (2018)

\bibitem{Soto2014}
R.~Soto, R.~Golestanian.
\newblock Run-and-tumble dynamics in a crowded environment: persistent
  exclusion process for swimmers.
\newblock \emph{Phys. Rev. E} \textbf{89}, 012706 (2014)

\bibitem{Slowman2016a}
A.~B. Slowman, M.~R. Evans, R.~A. Blythe.
\newblock Jamming and Attraction of Interacting Run-and-Tumble Random Walkers.
\newblock \emph{Phys. Rev. Lett.} \textbf{116}, 218101 (2016)

\bibitem{Slowman2017}
A.~B. Slowman, M.~R. Evans, R.~A. Blythe.
\newblock Exact solution of two interacting run-and-tumble random walkers with
  finite tumble duration.
\newblock \emph{J. Phys. A} \textbf{50}, 375601 (2017)

\bibitem{Demaerel2018}
T.~Demaerel, C.~Maes.
\newblock Active processes in one dimension.
\newblock \emph{Phys. Rev. E} \textbf{97}, 032604 (2018)

\bibitem{Malakar2018}
K.~Malakar, V.~Jemseena, A.~Kundu, K.~V. Kumar, S.~Sabhapandit, S.~N. Majumdar,
  S.~Redner, A.~Dhar.
\newblock Steady state, relaxation and first-passage properties of a
  run-and-tumble particle in one dimension.
\newblock \emph{J. Stat. Mech.} 043215 (2018)

\bibitem{Pietzonka2018}
P.~Pietzonka, U.~Seifert.
\newblock Entropy production of active particles and for particles in active
  baths.
\newblock \emph{J. Phys. A} \textbf{51}, {01LT01} (2018)

\bibitem{Karbach1998}
M.~Karbach, G.~M\"uller.
\newblock Introduction to the {Bethe} ansatz {I}.
\newblock \emph{arXiv:9809162}  (1998)

\bibitem{Diaconis2000}
P.~Diaconis, S.~Holmes, R.~M. Neal.
\newblock Analysis of a Nonreversible Markov Chain Sampler.
\newblock \emph{The Annals of Applied Probability} \textbf{10}, 3 (2000)

\bibitem{Kato1966}
T.~Kato.
\newblock \emph{Perturbation theory for linear operators}.
\newblock Springer (1966)

\bibitem{Heiss1990}
W.~D. Heiss, A.~L. Sannino.
\newblock Avoided level crossings and exceptional points.
\newblock \emph{J. Phys. A: Math Gen.} \textbf{23}, 1167 (1990)

\bibitem{vanKampen2007}
N.~G. van Kampen.
\newblock \emph{Stochastic Processes in Physics and Chemistry}.
\newblock Elsevier B.V., third edition (2007)

\bibitem{Heiss2012}
W.~D. Heiss.
\newblock The physics of exceptional points.
\newblock \emph{J. Phys. A: Math. Theor.} \textbf{45}, 444016 (2012)

\bibitem{Ryu2015}
J.~Ryu, W.~Son, D.~Hwang, S.~Lee, W.~Kim.
\newblock Exceptional points in couples dissipative dynamical systems.
\newblock \emph{Phys. Rev. E} \textbf{91}, 052910 (2015)

\bibitem{Kelly1979}
F.~P. Kelly.
\newblock \emph{Reversibility and Stochastic Networks}.
\newblock Wiley (1979)

\bibitem{Brattain2016}
E.~Brattain.
\newblock \emph{The Completeness of the Bethe Ansatz for the Asymmetric Simple
  Exclusion Process}.
\newblock Ph.D. thesis, University of California (2016)

\end{thebibliography}

\end{document}